\documentclass[%
twocolumn,
superscriptaddress,
nofootinbib,
longbibliography,
 amsmath,amssymb,
 aps,
pra
]{revtex4-1}

\usepackage{lipsum}
\usepackage[overload]{textcase}
\usepackage{mathbbol,amsmath,amsfonts,amssymb,bm,dsfont}
\usepackage{amsfonts}
\usepackage{dsfont}
\usepackage{graphicx,xcolor}
\usepackage{dcolumn}
\usepackage{bm}
\usepackage{hyperref}
\usepackage[mathlines]{lineno}
\usepackage{mathtools}
\usepackage{physics}
\usepackage[normalem]{ulem}
\usepackage{subfigure}
\usepackage{multirow}
\usepackage{siunitx}
\usepackage{amsthm}
\usepackage[title]{appendix}
\usepackage{comment}

\newcolumntype{x}[1]{>{\centering\hspace{0pt}}p{#1}}
\newcolumntype{L}[1]{>{\raggedright\arraybackslash}p{#1}}
\newcolumntype{C}[1]{>{\centering\arraybackslash}p{#1}}
\newcolumntype{R}[1]{>{\raggedleft\arraybackslash}p{#1}}

\providecommand{\ignore}[1]{}

\newif\ifcmnt
\cmnttrue
\ifdefined\cmntsoff\cmntfalse\fi

\newcommand{\cD}{{\cal D}}

\newcommand{\cH}{{\cal H}}

\newcommand{\cN}{{\cal N}}

\newcommand{\up}{{\uparrow}}
\newcommand{\down}{{\downarrow}}
\newcommand{\Up}{{\Uparrow}}
\newcommand{\Down}{{\Downarrow}}

\newtheorem*{theorem*}{Theorem}

\newcommand{\s}{\sigma}

\newcommand{\pih}{\frac{\pi}{2}}

\newcommand{\I}{{\mbox{\sc I}}}
\newcommand{\II}{{\mbox{\sc II}}}

\newcommand{\su}{\ket{\uparrow}}
\newcommand{\sd}{\ket{\downarrow}}
\newcommand{\E}[1]{\ket{E_{#1}}}

\newcommand{\ot}{\otimes}
\newcommand{\st}[1]{\ket{\psi_{#1}}}

\begin{document}

\title{An Operator Analysis of Contextuality Witness Measurements \\ for Multimode-Entangled Single Neutron Interferometry}

\author{Shufan Lu}
\affiliation{Department of Physics, Indiana University, Bloomington, IN
 47405,USA}
\affiliation{Indiana University Center for the Exploration of Energy and Matter, Bloomington, IN 47408, USA}
\affiliation{Indiana University Quantum Science and Engineering Center, Bloomington, IN 47408, USA}

\author{Abu Ashik Md.~Irfan}
 \affiliation{Department of Physics, Indiana University, Bloomington, IN
 47405,USA}
\affiliation{Indiana University Quantum Science and Engineering Center, Bloomington, IN 47408, USA}

\author{Jiazhou Shen}
\affiliation{Department of Physics, Indiana University, Bloomington, IN
 47405,USA}
\affiliation{Indiana University Center for the Exploration of Energy and Matter, Bloomington, IN 47408, USA}
\affiliation{Indiana University Quantum Science and Engineering Center, Bloomington, IN 47408, USA}

\author{Steve J. Kuhn}
\affiliation{Department of Physics, Indiana University, Bloomington, IN
 47405,USA}
\affiliation{Indiana University Center for the Exploration of Energy and Matter, Bloomington, IN 47408, USA}
\affiliation{Indiana University Quantum Science and Engineering Center, Bloomington, IN 47408, USA}
 
\author{W. Michael Snow}
\affiliation{Department of Physics, Indiana University, Bloomington, IN
 47405,USA}
\affiliation{Indiana University Center for the Exploration of Energy and Matter, Bloomington, IN 47408, USA}
\affiliation{Indiana University Quantum Science and Engineering Center, Bloomington, IN 47408, USA}
 
\author{David V. Baxter}
\affiliation{Department of Physics, Indiana University, Bloomington, IN
 47405,USA}
\affiliation{Indiana University Center for the Exploration of Energy and Matter, Bloomington, IN 47408, USA}
\affiliation{Indiana University Quantum Science and Engineering Center, Bloomington, IN 47408, USA}
 
\author{Roger Pynn}
\affiliation{Department of Physics, Indiana University, Bloomington, IN
 47405,USA}
\affiliation{Indiana University Center for the Exploration of Energy and Matter, Bloomington, IN 47408, USA}
\affiliation{Indiana University Quantum Science and Engineering Center, Bloomington, IN 47408, USA}
\affiliation{Neutron Sciences Directorate, Oak Ridge National Laboratory, Oak Ridge, TN, 37830}

\author{Gerardo Ortiz}
\affiliation{Department of Physics, Indiana University, Bloomington, IN
 47405,USA}
\affiliation{Indiana University Quantum Science and Engineering Center, Bloomington, IN 47408, USA}

\date{\today}

\begin{abstract}
We develop an operator-based description of two types of multimode-entangled single-neutron quantum optical devices: Wollaston prisms and radio-frequency spin flippers in inclined magnetic field gradients. This treatment is similar to the approach used in quantum optics, and is convenient for the analysis of quantum contextuality measurements in certain types of neutron interferometers. We describe operationally the way multimode-entangled
single-neutron states evolve in these devices, and provide expressions for the associated operators describing the dynamics, in the limit in which the neutron state space is approximated by a finite tensor product of distinguishable subsystems. We design entangled-neutron interferometers to measure entanglement witnesses for the Clauser, Horne, Shimony and Holt, and Mermin inequalities, and compare the theoretical predictions with recent experimental results. We present the generalization of these expressions to $n$ entangled distinguishable subsystems, which could become relevant in the future if it becomes possible to add neutron orbital angular momentum to the experimentally-accessible list of entangled modes. 
We view this work as a necessary first step towards a theoretical description of entangled neutron scattering from strongly entangled matter, and we explain why it should be possible to formulate a useful generalization of the usual Van Hove linear response theory for this case. We also briefly describe some other scientific extensions and applications which can benefit from interferometric measurements using the types of single-neutron multimode entanglement described by this analysis.   
\end{abstract}

\maketitle

\section{Introduction}\label{Sec: Introduction}


Conventional neutron interferometry, employing devices constructed from large perfect single crystals of silicon, has been used to explore various foundational aspects of quantum mechanics \cite{ni-book-2015}. Although neutron spin-echo devices are often viewed in terms of Larmor precession of the neutron spin, under special physical conditions, the action of these devices is more suitably interpreted in terms of neutron interferometry and quantum entanglement \cite{barnum-2004}. In this paper we describe entanglement properties and dynamics in single-neutron interferometry as realized with instruments typically utilized in spin-echo scattering angle measurement (SESAME) techniques \cite{Parnell-2015}. It is our belief that an entanglement-based formulation for neutron spin-echo spectroscopic techniques such as SESAME can enable the future development of a theory for single-particle multimode-entangled neutron scattering from condensed matter systems that will shed light on exotic, non-local correlations in materials. 

A spin-echo neutron interferometer \cite{plomp-thesis} employing either magnetic Wollaston prisms (MWPs) \cite{li-2014} or radio-frequency neutron spin flippers (RFNSFs) coupled with magnetic field boundaries inclined to the neutron momentum~\cite{parnell-2018} can entangle the spin, path (MWP), and energy states (RFNSF) of a single neutron \cite{barnum-2004}. We present a single-neutron quantum optics analysis to describe the results of recent entanglement witness \cite{mermin-1990} measurements realized in an experimentally-flexible neutron spin-echo setup~\cite{iu-2019}. Although similar entanglement witness measurements were conducted in the past using perfect crystal neutron interferometry \cite{hase-2003,hase-2010}, there are two main features of the experimental work conducted with our type of interferometer which are transformative. We generated and controlled single-particle entangled neutron states sensitive to spin-dependent interactions, with an entanglement length $\xi$ (distance between neutron paths) that can be adjustable from nanometers to microns, and an entanglement energy separation flexible as well from peV to sub neV with both two and three entangled distinguishable subsystems. Unlike the cm-scale entanglement lengths generated in the perfect crystal neutron interferometers, these much smaller length scales coincide with the typical range of length scales where one expects interesting entangled excitations in condensed matter systems to exist. Furthermore, we demonstrated the capability to continuously tune the number of entangled subsystems, or modes, from three to two by adjusting the entanglement energy separation, thereby ``turning off" the energy entanglement, which is one of the most relevant variables for coupling to condensed matter excitations. 
\ignore{
These properties and capabilities of this type of neutron spin-echo spectroscopy therefore lend themselves to new types of scattering investigations of entanglement of condensed matter using single-particle entangled neutrons as a probe. We believe that the same features that make traditional unentangled neutron scattering from condensed matter systems such a broadly-applicable technique are also very well-suited for quantitative investigations of entanglement in condensed matter, especially in the most physically-interesting cases of strongly-interacting many-body systems for which there is a need to unveil their emergent interactions.  
}



Naturally, the first step in a plan to develop entangled neutron scattering is to prove that the neutron states realized in the neutron spin-echo spectrometers of interest are in fact entangled. Quantum entanglement, or non-separability, is at the core of the fundamental difference between classical and quantum representations of reality \cite{bell-collection}. To test for the presence of entanglement, we perform a contextuality test. The Kochen-Specker theorem \cite{ks-1967,peres-1991, mermin-rmp-1993}  articulated and addressed the concept of quantum contextuality~\cite{peres-book}, which distills the essence of Bohr's complementarity concept by quantifying how the measured value of one quantum observable in a system depends on the results of other quantum observables being measured along with it in the same system~\cite{cabello-2019}.  For any set of quantum observables one can define contextuality witnesses which are expectation values of correlated observables chosen to be sensitive to the degree of entanglement of the state measured~\cite{chsh-1969,cs-1978,mermin-1990}. In a non-contextual theory, those witnesses satisfy certain (classical) bounds. If a measurement violates the relevant bounds it indicates the impossibility of non-contextual hidden variables (NCHV). 

In our setup we are able to entangle the spin, path (position or trajectory), and energy observables of a single neutron in either spin and path or in spin, path, and energy by using the action of 
a trapezoidal RFNSF.
The relevant bounds for these two cases are set by the Mermin inequality \cite{mermin-1990} for spin, path, and energy modes, and the Clauser, Horne, Shimony and Holt (CHSH) inequality \cite{chsh-1969,cs-1978} for spin and path subsystems alone. In the limit in which one can treat the relevant degrees of freedom of the neutron in terms of finite-dimensional subsystems, we theoretically demonstrate that as the separation $\hbar \Delta$ of the energy modes approaches zero, the Mermin contextuality inequality reduces to the CHSH one, as expected. Our experimental results \cite{iu-2019} are in good agreement with this expectation. We therefore claim that we have experimentally demonstrated a state-of-the-art quantum entangled single-neutron probe, and a neutron interferometer of unprecedented flexibility, which can be used to develop and explore new forms of entangled scattering experiments.

The validity of the approximation of the neutron state space  in terms of distinguishable subsystems,  which we employ in this paper, deserves a brief explanation. The subsystem (mode or qubit) treatment of the spin-$\frac{1}{2}$ degree of freedom of the neutron is complete as we can ignore the antineutron degrees of freedom present in principle in the 4-component neutron Dirac spinor in the extreme nonrelativistic limit we operate the experiment in. The subsystem treatment of the energy and path degrees of freedom, although an excellent approximation in both cases, is an approximation. For the neutron path degree of freedom, the validity of the distinguishable subsystem picture depends on having negligible overlap of the neutron coherence volumes associated with the two possible paths through the interferometer. As the neutron source itself possesses no coherence, the coherence volume is developed dynamically \cite{klein-1983} through the entanglement-free neutron interactions upstream of the apparatus which localizes the neutrons from the source well enough to form a beam with a well-defined average momentum direction. In our case the longitudinal and transverse coherence lengths are of tens of nm order \cite{KaiserCoherence1983} and from tens of nm to as large as \SI{80}{\micro\meter}, respectively, compared to the spin-echo separation of \SI{1.5}{\micro\meter} given \SI{0.4}{\nano\meter} wavelength neutrons. This number varies in different measurement methods and depends on whether it is a coherence volume defined by beam collimation or a wave front property \cite{TREIMER20061388}. For the neutron energy degree of freedom the validity of the finite-subsystem approximation depends on the sharpness of the radio-frequency (RF) field energy $\hbar \omega$ of the magnetic field which couples to the neutron magnetic moment compared to the energy separation 
between the two neutron energies in the static magnetic field inside the device. It also depends on a weak enough coupling between the neutron and the external RF magnetic fields, so that one can treat the interaction in terms of the exchange of one photon. All of the conditions stated above are very well satisfied in the experiments we conducted and in general in the instrumentation used in almost all similar neutron spin-echo spectrometers.   



The neutron interferometric setup we describe below~\cite{iu-2019} 
is well-suited for exploring spin dynamics in the most physically-interesting cases of highly-entangled many-body systems, for which there
is a need to develop new techniques that can unveil complex emergent behavior. For a quantitative interpretation of entangled neutron scattering one must also develop a scattering theory for entangled neutron beams on entangled systems. The description of the coherent interactions of the neutron with the macroscopic media, which generate the entanglement and enable the measurement of the relevant entanglement witnesses, can be viewed as the zero-order approximation to the theory of interest in the limit of elastic interactions with no entanglement. Such an entangled neutron beam can be thought of as a truly quantum probe of condensed matter. 

In addition to the three entangled neutron properties of spin, path, and energy whose various forms of entanglement we describe below, one can also imagine that in the future it may be possible to produce single-particle neutron states with additional orbital angular momentum (OAM) mode entanglement. If these efforts can be successful at the neutron single-particle level, one can imagine the future possibility of generating single-particle entangled neutron states in four different dynamical variables: spin, path, energy, and OAM. 
We therefore present the generalization of our formalism to cover some aspects of this general case as well.  This may lead to the development of new protocols for quantum estimation and metrology of fundamental physical constants~\cite{metrology-2007,qsensors-rep}.

This paper is organized as follows. In Sec. \ref{Sec: Background} we briefly review the concept of contextuality in measurements of physical 
properties, and the way it is quantified by means of entanglement witnesses. 
We next present a theoretical description of the devices utilized in the design of multimode-entangled single-neutron interferometers. In Sec. \ref{Sec: Entangler} we introduce two types of neutron devices that act as ``entanglers'' of the distinguishable properties of the neutron. They are the MWP and RFNSFs combined with field gradients inclined to the neutron momentum. We discuss the experimental setup for the entangled neutron interferometers in Sec. \ref{Sec: Interferometer}. The exact relation between the measured spin polarization of the neutron and the entanglement witnesses is explained in Sec. 
\ref{Sec: Statistical Analysis}. In Sec.~\ref{Generalization} we present the generalization to multiple-mode entanglement of $n$ distinguishable subsystems. 
In Sec. \ref{Experimental Results} the recent experimental results are presented and analyzed in light of the theory developed.
Section \ref{Discussion and Outlook} discusses ideas for further development of a broadly-applicable theory for entangled neutron scattering from entangled systems in the same linear-response limit where the usual Van Hove treatment of unentangled neutron scattering holds. Furthermore, we also provide an outlook for future extensions to add OAM and perform quantum-enhanced metrology of fundamental physical constants \cite{metrology-2007}. 
Finally, the Appendix applies the operator formalism developed in this paper to the first triply-entangled neutron contextuality test conducted using a perfect crystal neutron interferometer \cite{hase-2010}.   

\section{Background}
\label{Sec: Background}

\subsection{Testing Quantum Contextuality: \\ CHSH Inequality}\label{Sec: CHSH}

A fundamental characteristic of a quantum description of physical phenomena is its contextual nature.  
Measurement outcomes of compatible sets of quantum observables, known as contexts, cannot reveal pre-existent 
values of the properties measured. The measured values depend upon the context. 
The Kochen-Specker theorem proves that NCHV theories
cannot reproduce the empirical predictions of quantum physics~\cite{ks-1967,peres-1991, mermin-rmp-1993}. 
It asserts that in a Hilbert space of dimension larger or equal to three, it is impossible to
associate determinate probabilities, $p_i= 0$ or $1$, with every
projection operator~$P_i$, in such a way that, if a set of commuting
$P_i$ satisfies $\sum_i P_i = \mathbb{1}$, then $\sum_i p_i = 1$. Bell non-locality 
~\cite{Bell_1964,Bell_1966,Scaranibook}, on the other hand, refers to the fact that the measurement outcomes of 
spacelike separated observables are not independent, and therefore, cannot be reproduced under the 
assumption of local realism proposed by Einstein, Podolsky, and Rosen (EPR) \cite{epr-1935}. 
Unlike Bell non-locality, quantum contextuality is independent of the spacetime structure of the measurements 
and, therefore, experiments to test it need not be performed involving spacelike separated events.

Although quantum contextuality and Bell non-locality represent independent concepts \cite{cabello-2019}, one can associate~\cite{peres-book} every Bell inequality to a quantum contextuality inequality using Neumark's dilation argument \cite{Peres1990-2}, where the non-contextual bound for the contextuality inequality equals the local bound 
of the Bell inequality. Then, maximum violation of the contextuality inequality will correspond to the maximum violation of the Bell inequality predicted by quantum mechanics. In this paper, we adapt a particular Bell inequality, namely CHSH (Clauser, Horne, Shimony, and Holt) inequality~\cite{chsh-1969} 
to test quantum contextuality of a neutron state in a particular experimental set-up. We use such a contextuality test to prove that our neutron beam is entangled. 

We treat spin ($s$) and path ($p$) degrees of freedom as two distinguishable subsytems~\cite{barnum-2004,cabello-2008}, and associate to our system the tensor product Hilbert state space $\mathcal{H}=\mathcal{H}_{s}\otimes\mathcal{H}_{p}$.
 Both $\cH_s$ and $\cH_p$ describe two-dimensional (qubit) subsystems: $\cH_s$ is the usual subspace of a nonrelativistic two-component spin-$\frac{1}{2}$ spinor and $\cH_p$ is the subspace spanned by two different path states describing the neutron's trajectory, which are far enough apart that one can neglect their spatial overlap. We define two pairs of observables:  $\sigma^s_{u(\alpha_i)}$ and $\sigma^p_{v(\chi_j)}$ acting on the corresponding subsystems, with $i,j\in\{1,2\}$, and $u(\alpha), v(\chi)$ labeling operators associated with angles $\alpha$ and $\chi$ in the $x$-$y$ plane of the corresponding Bloch spheres
\begin{eqnarray}
    \s_{u(\alpha)}^{s}&=&\cos{\alpha}\, \s^{s}_x+\sin{\alpha}\, \s^s_y     \label{Eq: Def_Pauli_spin}\\
    \s_{v(\chi)}^{p}&=&\cos{\chi}\, \s^{p}_x+\sin{\chi}\, \s^p_y.  \label{Eq: Def_Pauli_path} 
\end{eqnarray}
Having introduced the observables, we now define the CHSH witness $S$ 
\begin{equation} \label{Eq: CHSH Witness}
S=E(\alpha_1, \chi_1)+E(\alpha_1, \chi_2)+E(\alpha_2, \chi_1)-E(\alpha_2, \chi_2) .
\end{equation}
where $E(\alpha, \chi)$ represent the expectation value of $\s^s_{u(\alpha)}\s_{v(\chi)}^p$ over a state $\ket{\Psi} \in {\cal H}$, i.e. $E(\alpha, \chi)=E \left[ \s^s_{u(\alpha)}\s_{v(\chi)}^p \right]=
\bra{\Psi} \s^s_{u(\alpha)}\s_{v(\chi)}^p \ket{\Psi}$. No classical assignments of eigenvalues of observables by a local hidden variable theory can violate the CHSH inequality 
\begin{eqnarray}
|S|\leqslant 2 ,
\label{CHSHineq}
\end{eqnarray}
but quantum mechanical expectations can. The maximum value for $S$ set by quantum mechanics is the Tsirelson bound
$2 \sqrt{2}$ \cite{Scaranibook},
\begin{eqnarray}
    -2\leqslant &S& \leqslant 2 \ \ \ \ \ \ \ \ \ \   \mbox{(classical statistics)}   \nonumber\\
    -2\sqrt2\leqslant & S & \leqslant2\sqrt2  \ \ \ \ \ \  \mbox{(quantum statistics)}.    \nonumber
\end{eqnarray}
Any state violating the CHSH inequality \eqref{CHSHineq} is necessarily an entangled state in the spin and path degrees of freedom.

\subsection{Mermin Contextual Inequality}\label{Sec: Mermin}

In 1990, Mermin proposed a stronger version of the Bell inequality~\cite{mermin-1990-qmr,mermin-1990}, now called the Mermin inequality in the quantum information literature. He considered correlated measurements on entangled quantum states with $n \geqslant3$ subsystems, and showed that the size of the violation of his proposed inequality in quantum mechanics increases exponentially with $n$. 
We apply Mermin's inequality to test quantum contextuality of the single-neutron measurements, which as mentioned above does not require measurements to be spacelike separated. We consider the spin, path and energy degrees of freedom of the neutron as three distinguishable subsystems. The Hilbert space describing such a system can be expressed in terms of a tensor product $\mathcal{H}=\mathcal{H}_{s}\otimes\mathcal{H}_{p}\otimes\mathcal{H}_{e}$, where $\mathcal{H}_{s}$, $\mathcal{H}_{p}$ and $\mathcal{H}_{e}$ are the spin, path and energy subspaces. All three subspaces are two dimensional:  $\cH_s$ is the usual subspace of a nonrelativistic two-component spin-$\frac{1}{2}$ spinor, while $\cH_p$ ($\cH_e$) is a subspace spanned by two different path states (energy states) of neutron's trajectory (energy). For this system we write the Mermin witness as
\begin{equation} \label{Eq: Mermin-Witness}
   M=E[\s^{s}_{x}\s^{p}_{x}\s^{e}_{x}]-E[\s^{s}_{x}\s^{p}_{y}\s^{e}_{y}]-E[\s^{s}_{y}\s^{p}_{x}\s^{e}_{y}]-E[\s^{s}_{y}\s^{p}_{y}\s^{e}_{x}], 
\end{equation}
where $E[\s^{s}_{x,y}\s^{p}_{x,y}\s^{e}_{x,y}]$ is the expectation value of $\s^{s}_{x,y}\s^{p}_{x,y}\s^{e}_{x,y}$ over a state $\ket{\Psi} \in {\cal H}$. No classical assignments of eigenvalues of these observables by a local hidden variable theory can violate the Mermin inequality 
\begin{eqnarray}
|M|\leqslant 2 ,
\label{Merminineq}
\end{eqnarray}
while quantum mechanical expectations may. The maximum value for $M$ set by quantum mechanics is $4$, 
\begin{eqnarray}
    -2\leqslant &M&\leqslant2 \ \ \ \ \ \mbox{(classical statistics)}     \nonumber\\
    -4\leqslant &M&\leqslant4  \ \ \ \ \mbox{(quantum statistics)}.    \nonumber
\end{eqnarray}
Any state violating the Mermin inequality \eqref{Merminineq} is necessarily an entangled state in spin, path and energy degrees of freedom.

\section{neutron entangler devices}\label{Sec: Entangler}

An essential part of any quantum interferometer is the entangler which generates the desired entangled state starting from an unentangled initial state. In this section we discuss the construction and working principles of two specific entangler devices for a single-neutron state, i) a pair of MWPs and ii) a pair of RFNSFs, with magnetic field boundaries inclined with respect to the neutron momentum, used in the interferometer discussed in Sec. \ref{Sec: Interferometer}. The first one entangles the incident neutron in spin and path modes, while the second one can entangle the neutron either into (two) spin and path modes or into three modes of spin, path and energy.

\subsection{Magnetic Wollaston Prism: \\ \ \ \ \ Polarizing Beam Splitter}\label{Sec: MWP construction}

\begin{figure}[htb]
    \centering
    \includegraphics[scale=0.40]{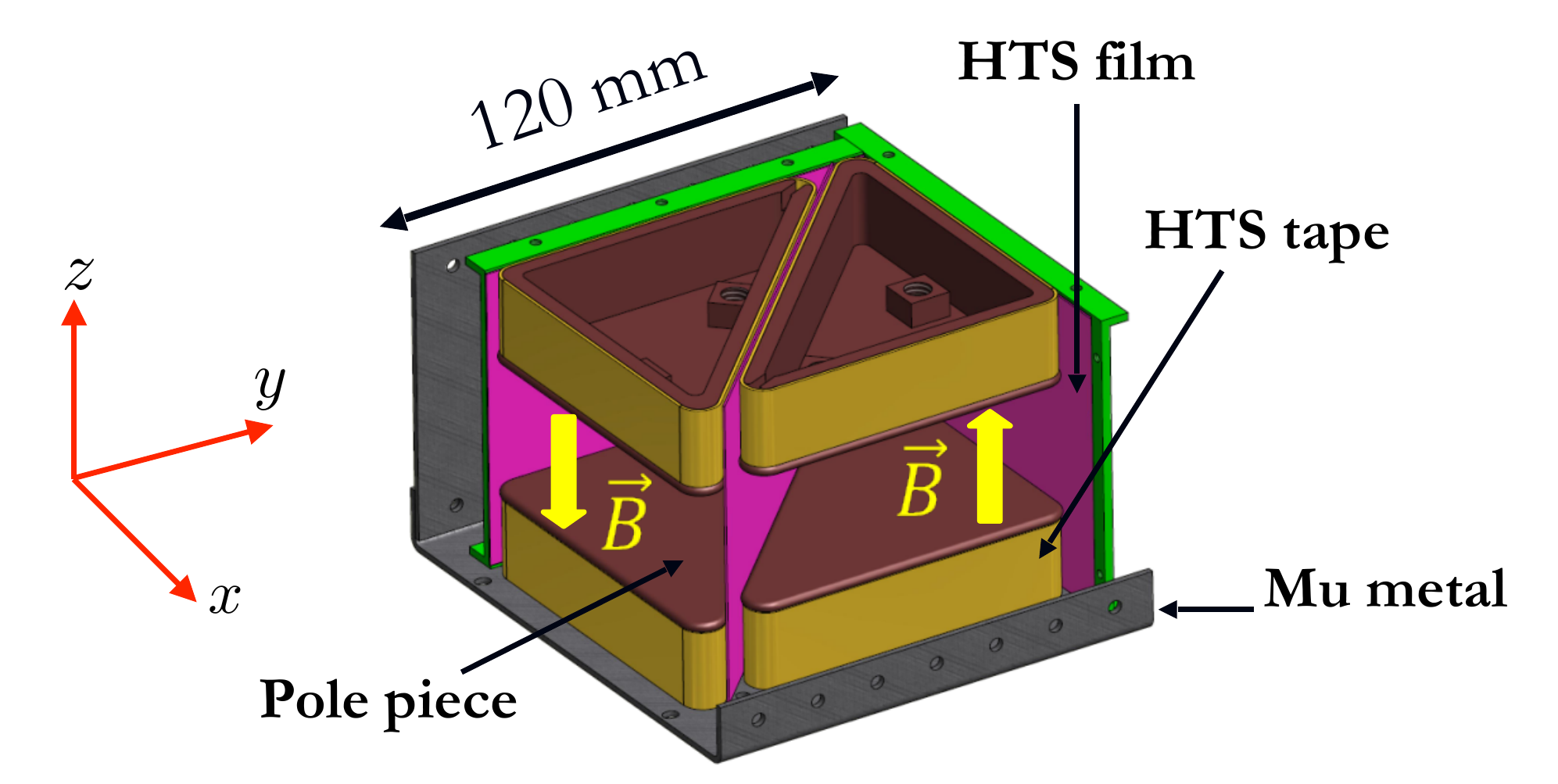}
    \hspace*{-0.5cm} {\includegraphics[scale=1.20]{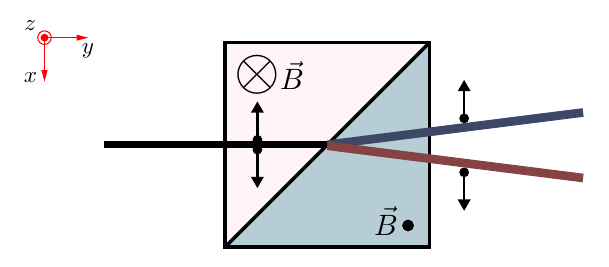}}
    \caption{The top part shows a detailed view of a MWP. Each triangular shaped region has a magnetic field $\vec B$ anti-parallel to the other one. The geometry of this arrangement promotes efficient magnetic flux return and helps to reduce unwanted neutron optical aberrations in the final neutron spin state from stray magnetic fields. The gold color highlights the high-$T_{c}$ superconducting (S) tape which, in combination with the high-$T_{c}$ S film coated onto the sapphire substrates that the neutron beam passes through, is important for the creation of a spatially-sharp magnetic field boundary along the hypotenuse of each triangle. The bottom part shows the idealization employed to work out the dynamics of the entanglement generation in terms of a refraction of the incident neutron state by the optical potential of the magnetic field created by the MWP. As a static magnetic field is a birefringent medium for a neutron, the two neutron spin states are refracted into two different directions as shown (spin arrows point along the $z$-direction). However, as the potential energy from this MWP is time-independent there is no energy exchange between the neutron and the device and so the final kinetic energy of the neutron as it exits the device is unchanged.}\label{fig:animals} \label{MWP: structure}
\end{figure}

In this section we show how a pair of MWPs work together to entangle a neutron beam in spin and path subsystems or modes. Just like an optical Wollaston prism refracts the two polarization components of incident light into different directions~\cite{Hecht2002}, a MWP~\cite{li-2014} coherently refracts the two spin components of the incident neutron beam into two different directions. The device is cubic in shape and divided into two right-angled triangles as shown in Fig.~\ref{MWP: structure}. Each triangular region has a pair of superconducting coils in both the upper and lower faces. The large currents in the superconductors create strong static antiparallel magnetic fields $\vec B$ of equal magnitude in both triangular regions which the neutrons pass directly through with negligible decoherence. The triangular magnetic field geometry is sharply defined by the Meissner effect from high-$T_c$ films on all three sides. Neutrons which pass quickly enough through the sharply-defined magnetic field discontinuity at the interface between these two triangular regions experience a nonadiabatic change in their potential energy from the $-\vec{\mu} \cdot \vec{B}$ interaction of the neutron magnetic moment $\vec{\mu}$ with the  field. The motion normal to the magnetic field boundary is well modeled as a one-dimensional potential energy step whose sign is different for the two neutron spin states, and the motion parallel to the magnetic field boundary sees no gradient. The neutron therefore refracts from this step change in the potential with an amplitude that can be calculated very simply using one-dimensional quantum mechanics of a single particle. The final energies and momenta of the two refracted components of the initial neutron state are determined by applying energy and momentum conservation at the boundary. One spin projection along $\vec{B}$ gains kinetic energy while the other loses kinetic energy, and so these two neutron spin states exit the device in different directions. By the linearity of quantum mechanics, an incident neutron state which is a coherent superposition of these two spin components becomes entangled in spin and momentum by this first component of the MWP \cite{Felcher1995} as the two different momenta have the same magnitude but are traveling in different directions. If the incoming neutron spin direction is normal to the magnetic field directions inside the MWP, the two refracted amplitudes have the same magnitude. The field inside the MWP is chosen to be much stronger than the guide fields that are applied throughout the apparatus to minimize possible decoherence in the rest of the interferometer during the passage of the neutron between the optical elements. In this paper we will ignore the very small perturbing effects of the guide fields, but if needed they could be taken into account to help quantify possible dephasing effects in sensitive contextuality measurements. 

\begin{figure}[htb]
    \includegraphics[scale=0.8]{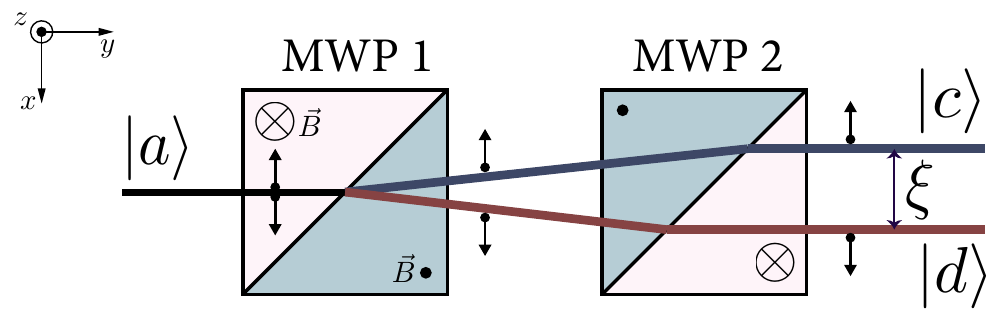}
    \caption{A pair of MWPs entangle the incident neutron state in spin and path degrees of freedom. Each MWP consists of two regions with antiparallel magnetic fields $\vec B$: the direction of the magnetic field in the grey shaded regions is in the $+z$-direction, while the direction of the magnetic field in unshaded regions is $-z$-direction. A neutron beam incident in path $a$ with $z$-spin component up refracts to path $c$ while its down spin component refracts to path $d$. The entanglement length, $\xi$, is defined as the separation between neutron paths.}
    \label{Entagnler: MWP}
\end{figure}

As shown in Fig.~\ref{Entagnler: MWP}, two oppositely-oriented but otherwise identical MWPs acting in sequence entangle a polarized incident unentangled neutron state (normal to the internal magnetic field direction) into spin and path modes. Since the neutron is a spin-$\frac{1}{2}$ particle with only two magnetic states, the dimensionality of the path subspace is also two. For our analysis we consider the idealized limiting case of a neutron beam incident perpendicular to the front and back faces of the MWP and with the spin state in the $x$-$y$ plane. The first component of the MWP entangles the neutron beam into spin and momentum as described above while the second MWP, which has opposite magnetic field orientations to that of the first MWP, makes the two incident spin trajectories parallel so that the magnitude and direction of the momenta are the same, but the spatial wave functions do not overlap and are separated by the entanglement length $\xi$, thus entangling the exiting neutron state into spin and path modes as shown in Fig.~\ref{Entagnler: MWP}. We denote the incident path as $a$, and the two outgoing paths as $c$ (spin-up) and $d$ (spin-down). The transition amplitudes can be read directly from the map
\begin{eqnarray}
    \ket{\up \, a} &\mapsto& \ket{\up \, c},     \label{Eq: transition_up_MWP}\\
    \ket{\down \, a} &\mapsto& \ket{\down \, d}.     \label{Eq: transition_down_MWP}
\end{eqnarray}
\begin{figure}[htb]
    \includegraphics[scale=0.23]{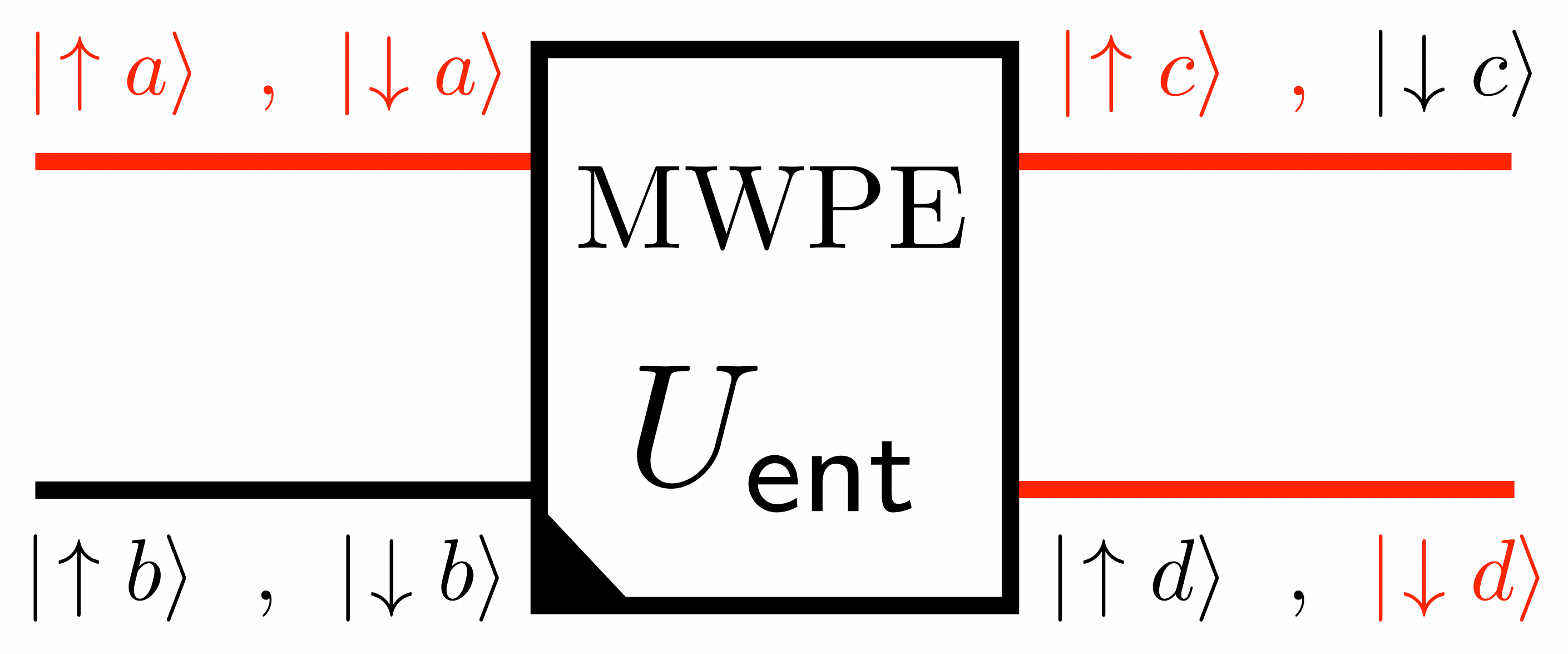}
    \caption{Unitary operator, $U_{\sf ent}$,  associated to the MWP entangler (MWPE) with all the incoming and outgoing spin/paths indicated. Relevant transition amplitudes are marked in red.}
    \label{Fig: Unitary: MWP}
\end{figure}

We derive the unitary operator corresponding to this MWP entangler (MWPE). 
First we derive an operator $U_{\sf BS}$ which accounts for the beam-splitting operation of the MWPE. We label the transition amplitudes from path $a$ to path $c$ as ($t_{\up, \down}$) and the transition amplitudes from path $a$ to path $d$ as ($r_{\up, \down}$). We label $b$ as the incident path transmitted to $d$ and reflected to $c$ as shown them in Fig.~\ref{Fig: Unitary: MWP}. In our interferometer we only used the $\ket{a}$ input state. We express $U_{\sf BS}$ in the bases $  \left\{\{\su,\sd\}\ot\{\ket{a},\ket{b}\}\right\} \mapsto  \left\{\{\su,\sd\}\ot\{\ket{c},\ket{d}\}\right\}$ as
\begin{eqnarray} \nonumber
    U_{\sf BS}&=&
    \left(
    \begin{array}{cc}
        1   & 0\\
        0 & 0\\
    \end{array}
    \right)\otimes 
    \left(
    \begin{array}{cc}
        t_\up   & i r_\up\\
        i r_\up & t_\up\\
    \end{array}
    \right)+
    \left(
    \begin{array}{cc}
        0   & 0\\
        0 & 1\\
    \end{array}
    \right)\otimes 
    \left(
    \begin{array}{cc}
        t_\down   & i r_\down\\
        i r_\down & t_\down\\
    \end{array}
    \right)\\
    &=&\left(
    \begin{array}{cccc}
    t_\up   & i r_\up & 0          & 0 \\
    i r_\up &  t_\up  & 0          & 0 \\
    0       & 0       & t_\down    & i r_\down \\
    0       & 0       & i r_\down  & t_\down\\
    \end{array}
    \right) , 
\end{eqnarray}
with $t_{\up,\down}, r_{\up,\down} \in {\mathbb R}$ and  $|t_{\up,\down}|^2+|r_{\up,\down}|^2=1$. Note that the first basis vector for each subsystem corresponds to the vector $\binom{1}{0}$, while the second one to $\binom{0}{1}$. The transition amplitudes are described in Eq. \eqref{Eq: transition_up_MWP} and \eqref{Eq: transition_down_MWP} and we set $t_\up=1=r_\down$. Now we consider the Larmor precession \cite{Rekveldt2002} that is produced from the static magnetic fields of the MWPs. By construction all the magnetic fields in the entangler are either parallel or antiparallel to the $z$-axis, and the symmetric construction minimizes the size of the magnetic field components experienced by the neutrons in the other directions. Therefore, we can combine effects from all the magnetic fields into a single unitary operator  $\cD(\hat{z},-2\varphi)=\exp[i\varphi \sigma^s_{z}]$. Using $U_{\sf ent}=\cD(\hat{z},-2\varphi) \, U_{\sf BS}$ one can derive 
\begin{equation}\nonumber
    U_{\sf ent}=\left(
     \begin{array}{cccc}
        e^{i\varphi} & 0 & 0 & 0 \\
        0  & e^{i\varphi} & 0 & 0 \\
        0 & 0 & 0 & i e^{-i\varphi}  \\
        0 & 0 & ie^{-i\varphi} & 0\\
    \end{array}
    \right).
\end{equation}
This matrix will introduce complex phases in the transition amplitudes defined in Eq. \eqref{Eq: transition_up_MWP} and \eqref{Eq: transition_down_MWP}. 
If the incident state is $\frac{\ket{\up}+\ket{\down }}{\sqrt 2}\ot \ket{a}$, then this entangler creates a Bell state
\begin{equation} \label{Eq: Bell state prep_MWP}
    U_{\sf ent} \frac{\ket{\up }+\ket{\down}}{\sqrt 2}\ot  \ket{a}=  e^{i \varphi}\frac{\ket{\up \, c }+\ket{\down \, d }}{\sqrt 2} ,
\end{equation}
where we have redefined the basis states $\ket{a}\rightarrow  \ket{a}$, $ \ket{c}\rightarrow  \ket{c}$, $i e^{-i 2 \varphi} \ket{b}\rightarrow  \ket{b}$, and $i e^{-i 2 \varphi} \ket{d}\rightarrow  \ket{d}$, to get rid of the relative phase between $\ket{\up \, c }$ and $\ket{\down \, d }$. 
In this new basis
\begin{equation}\nonumber
    U_{\sf ent}=\left(
     \begin{array}{cccc}
        e^{i \varphi} & 0 & 0 & 0 \\
        0  & e^{i  \varphi} & 0 & 0 \\
        0 & 0 & 0 & -e^{-i 3 \varphi}  \\
        0 & 0 & e^{i \varphi} & 0\\
    \end{array}
    \right).
\end{equation}

\subsection{RF Flipper : Energy Entangler}\label{Sec: RF-flipper construction}

\ignore{
\begin{figure}[htb]
    \includegraphics[scale=0.12]{RFfg-photo.jpg}
    \caption{A picture from the ISIS experimental set-up~\cite{iu-2019}. The neutron beam travels from left to right in the picture. The two RFNSFs inside the blue rectangle in the picture can create entanglement in the neutron state in energy modes as described in the text.}
    \label{Fig: RF_Instrument}
\end{figure}
}

In this section we show how a pair of RFNSFs combined with static magnetic field boundaries inclined relative to the neutron momentum can work together to entangle an incident neutron state in its different degrees of freedom. This RFNSF group (RFNSFG) entangler can operate in two modes: it entangles neutrons either in the two degrees of freedom of spin and path or in the three degrees of freedom of spin, path, and energy. As shown below, the spin, path, and energy entanglement can be continuously tuned into spin and path entanglement by adjusting the RF magnetic field frequency. 

\begin{figure}[htb]
    \includegraphics[scale=0.45]{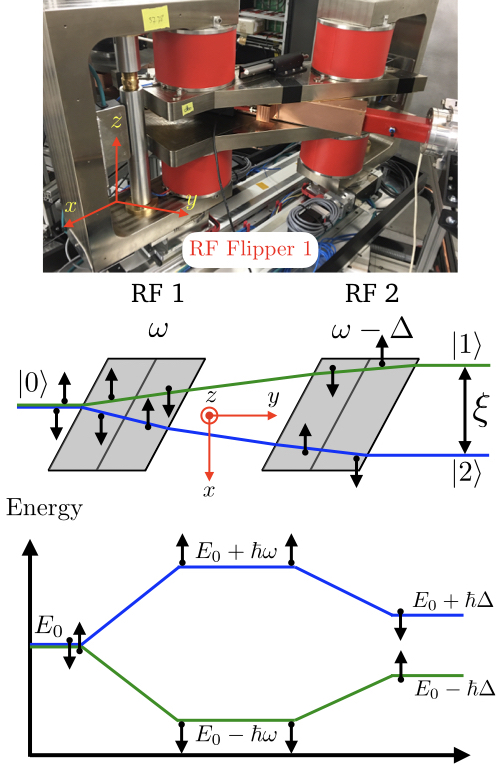}
    \caption{The top part shows a picture of a single RFNSF in the ISIS experimental set-up~\cite{iu-2019}. The 
    middle panel displays a pair of RFNSFs combined, i.e., a RFNSFG, that works as an entangler, with static magnetic field boundaries inclined relative to the $y$-direction. The first RFNSF (RF 1) combined with refraction at the inclined field boundary entangles the neutron into spin (spin arrows point along the $z$-direction), path and energy modes. The second RF 2 can either change the entanglement of the spin, path and energy modes or output only a spin and path entangled subsystems when $\Delta=0$.  The bottom part of the figure shows the kinetic energies $E_{\pm}=E_{0}\pm \hbar \Delta$ of the two spin components. For $\Delta=0$, both spin components have the same energy $E_0$ upon exiting the device.}
    \label{Fig: RF-Working principle}
\end{figure}

\begin{figure}[htb]
    \includegraphics[scale=0.22]{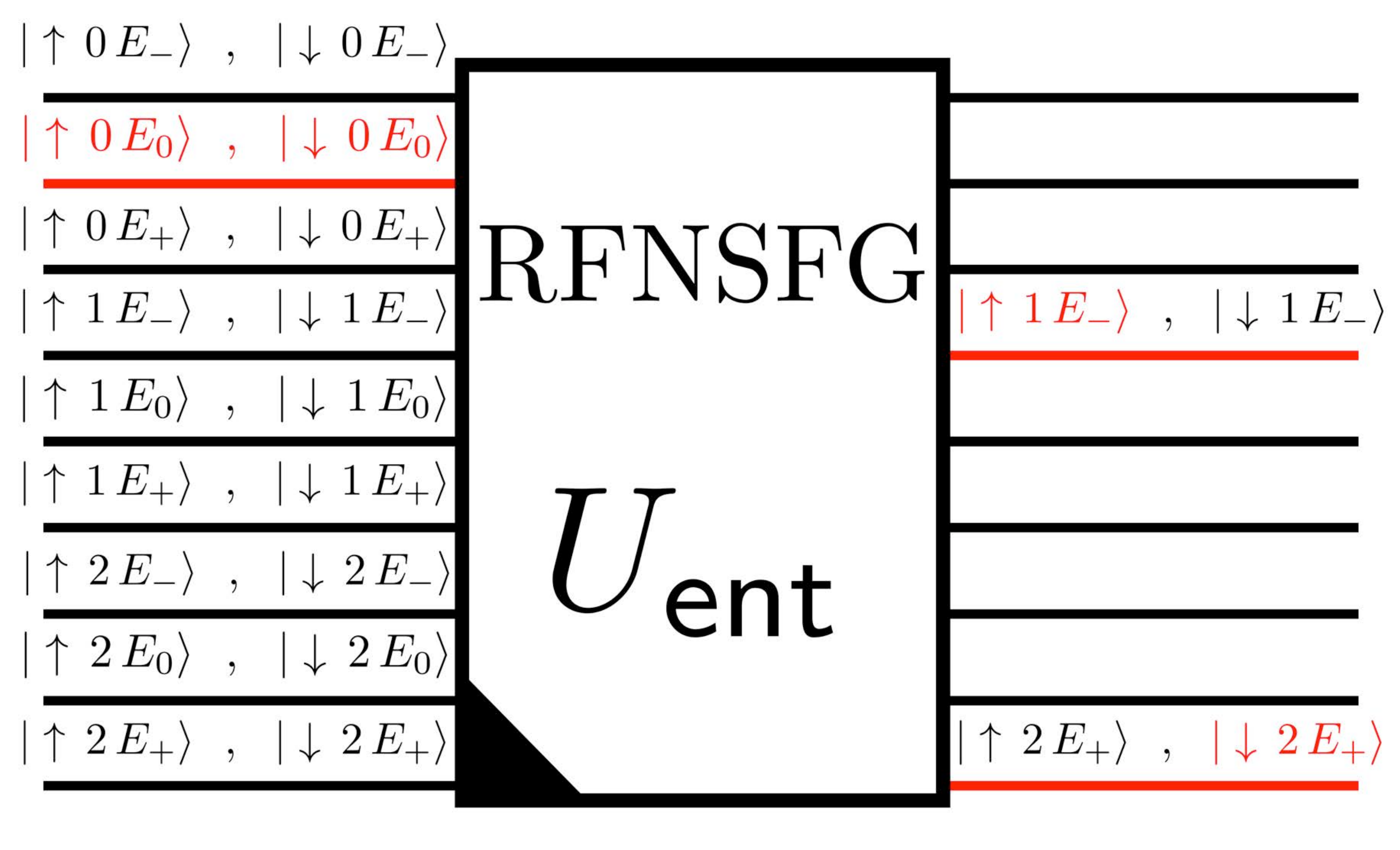}
    \caption{The entangler constructed from the RFNSFG set to generate three-mode entanglement can be expressed as a unitary operator ($U_{\sf ent}$). The  transition amplitudes marked in red correspond to $\ket{\up\, 0 \, E_0}\mapsto \ket{\up \, 1 \, E_-}$   and $\ket{\down\, 0 \, E_0}\mapsto \ket{\down \, 2 \, E_+}$.}
    \label{Fig: RF(3)-Unitary}
\end{figure}

We denote the neutron path state incident upon the entangler with energy $E_0$ as $\ket{0}$. The entangler consists of two RFNSFs (RF 1 and RF 2) and a static magnetic field with two parallel inclined magnetic field boundaries as shown in Fig.~\ref{Fig: RF-Working principle}. The static field acts in both parallelogram regions. The inclined interface between the static magnetic field boundaries generates a large magnetic field gradient with a component normal to the neutron beam momentum that refracts the incoming neutron beam as discussed above for the MWPs. The RFNSFs with the field gradients act on the neutron spin and path degrees of freedom in the same way as described above for the MWPs. An external RF field of frequency $\omega$ and amplitude chosen so that on average only one RF photon is exchanged between the neutron and the field flips the spin of the neutron \cite{Maruyama-2003}. The spin up component loses energy and the spin down component gains energy under this photon exchange. The second RFNSF operates at a different frequency $\omega-\Delta$.  The neutron spin flips again by exchanging a photon of energy $\hbar(\omega-\Delta)$ so that the spin up component has kinetic energy $E_-=E_0-\hbar\Delta$ and the spin down component has kinetic energy $E_+=E_0+\hbar\Delta$. The final state of the neutron is therefore entangled in spin, path, and energy degrees of freedom. Denoting the two outgoing paths as $1$ (spin-up) and $2$ (spin-down) with entanglement length $\xi$, the transition amplitudes become 
(see Fig. \ref{Fig: RF(3)-Unitary})
\begin{eqnarray}
    \ket{\up\, 0 \, E_0}&\mapsto& \ket{\up \, 1 \, E_-}   \label{Eq: transition_up_RF3}\\
    \ket{\down\, 0 \, E_0}&\mapsto& \ket{\down \, 2 \, E_+}.    \label{Eq: transition_down_RF3}
\end{eqnarray}

As for the MWPs, a small static magnetic field $\vec B_0$ is applied over the whole interferometer to minimize the generation of decoherence of the neutron spin from the environment. We neglect the effect of this small field in the entangler. 

To derive the unitary operator for this entangler, we write down the Hamiltonian $H$ which implements the transition amplitudes in Eq. \eqref{Eq: transition_up_RF3} and \eqref{Eq: transition_down_RF3} as
\begin{equation}\nonumber
    H=(\ket{\up \, 1}\bra{\up \, 0}   +\ket{\down \, 0}\bra{\down \, 2})\ot T+\mbox{h.c.},
\end{equation} 
where $T=\E{0}\bra{E_+}+\E{-}\bra{E_0}$, and $\mbox{h.c.}$ stands for Hermitian conjugate. The corresponding propagator is $\exp[\frac{-i H t}{\hbar}]$. By expanding the exponential one can show  
\begin{equation}\nonumber
    \exp[\frac{-i H t}{\hbar}]=\mathbb{1} -i\sin \frac{t}{\hbar} \, H -\left(1-\cos\frac{t}{\hbar}\right)H^2,
\end{equation}
where we used $H^3=H$. For $t=\frac{\pi \hbar}{2}$, 
\begin{equation}    \nonumber
    \exp[\frac{-i \pi H}{2}]= \mathds{1} - i (V_\uparrow+V_\downarrow) ,
\end{equation}
with ($\dagger$ denotes the Hermitian conjugate)
\begin{eqnarray}
    V_\uparrow &=& P_\up \ot \big(\ket{1} \bra{0}\ot T + \ket{0} \bra{1}\ot T^\dagger -iP^{\uparrow}_{pe} \big) \nonumber \\ 
    V_\downarrow&=& P_\down \ot \big(\ket{0} \bra{2}\ot T +\ket{2} \bra{0}\ot T^\dagger - i P^{\downarrow}_{pe}\big). \nonumber 
\end{eqnarray}
Here, $P_\psi=\ket{\psi}\bra{\psi}$ represents the projector onto the state $\ket{\psi}$, $P^{\uparrow}_{pe}=P_{0 E_0}+P_{0 E_+}+P_{1 E_-}+P_{1 E_0}$ and $P^{\downarrow}_{pe}=P_{0 E_-}+P_{0 E_0}+P_{2 E_0}+P_{2 E_+}$. This propagator includes the transitions discussed in Eq. \eqref{Eq: transition_up_RF3} and \eqref{Eq: transition_down_RF3}. The time $t$ is determined by the incident neutron speed through the device as both $\hbar \Delta \ll E_{0}$ and $\hbar \omega \ll E_{0}$. We also need to include the Larmor precession phases. Let the effects from all the magnetic fields be combined into a single unitary matrix $\cD(\hat{z},-2\varphi)=\exp[i\varphi \sigma^s_{z}]$ as for the MWPs. Using $U_{\sf ent}=\cD(\hat{z},-2\varphi)\exp[\frac{-i \pi H }{2}]$, one can show that
\begin{equation} \nonumber
    U_{\sf ent} = e^{i\varphi}(P_{\uparrow}-i V_\uparrow)+e^{-i\varphi}(P_{\downarrow}- iV_\downarrow).  
\end{equation}
This matrix possesses complex phases in the transition amplitudes as discussed in Eqs. \eqref{Eq: transition_up_RF3} and \eqref{Eq: transition_down_RF3}. As before we redefine the basis: $-i e^{i\varphi}\ket{1}\rightarrow \ket{1}$ and $-i e^{-i\varphi}\ket{2}\rightarrow \ket{2}$ to get
\begin{multline}
    U_{\sf ent}=
    e^{i\varphi}P_{\uparrow}+P_\up \otimes \big( \ket{1} \bra{0}\ot T -e^{i 2 \varphi} \ket{0} \bra{1}\ot T^\dagger-e^{i\varphi} P^\uparrow_{pe}\big)   \nonumber\\
   +e^{-i\varphi}P_{\downarrow}+P_\down \otimes \big(\ket{2} \bra{0}\ot T^\dagger  -e^{-i 2 \varphi}\ket{0} \bra{2}\ot T -e^{-i\varphi} P^\downarrow_{pe}\big).   \nonumber
\end{multline}
If the incident beam is $\frac{\ket{\up }+\ket{\down}}{\sqrt 2}\otimes\ket{0 \, E_0}$ then this entangler creates the GHZ state 
\begin{equation} \label{Eq: GHZ state prep_RF}
    U_{\sf ent} \frac{\ket{\up }+\ket{\down}}{\sqrt 2}\ot \ket{0 \, E_0}=  \frac{\ket{\up \,  1  \, E_-}+\ket{\down  \,  2  \, E_+}}{\sqrt 2}.
\end{equation}

\begin{figure}[htb]
    \includegraphics[scale=0.20]{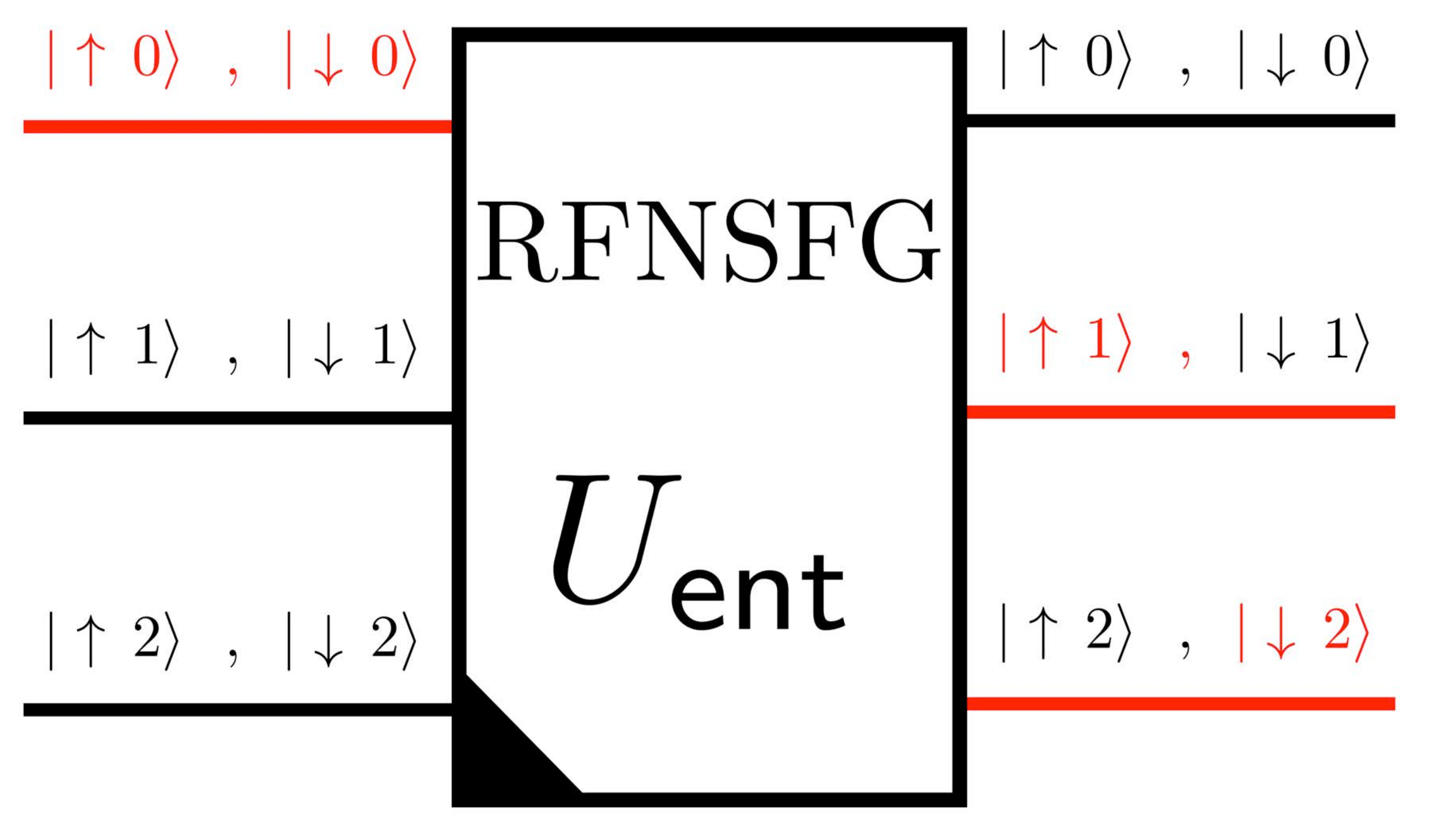}
    \caption{Entangler ($U_{\sf ent}$) constructed from the RFNSFG set to generate two-mode entanglement. Transition amplitudes marked in red correspond to  $\ket{\up\, 0}\mapsto \ket{\up \, 1 }$   and $\ket{\down\, 0}\mapsto \ket{\down \, 2}$.}
    \label{Fig: RF(2)-Unitary}
\end{figure}


\begin{figure*}[htb]
    \includegraphics[scale=0.6]{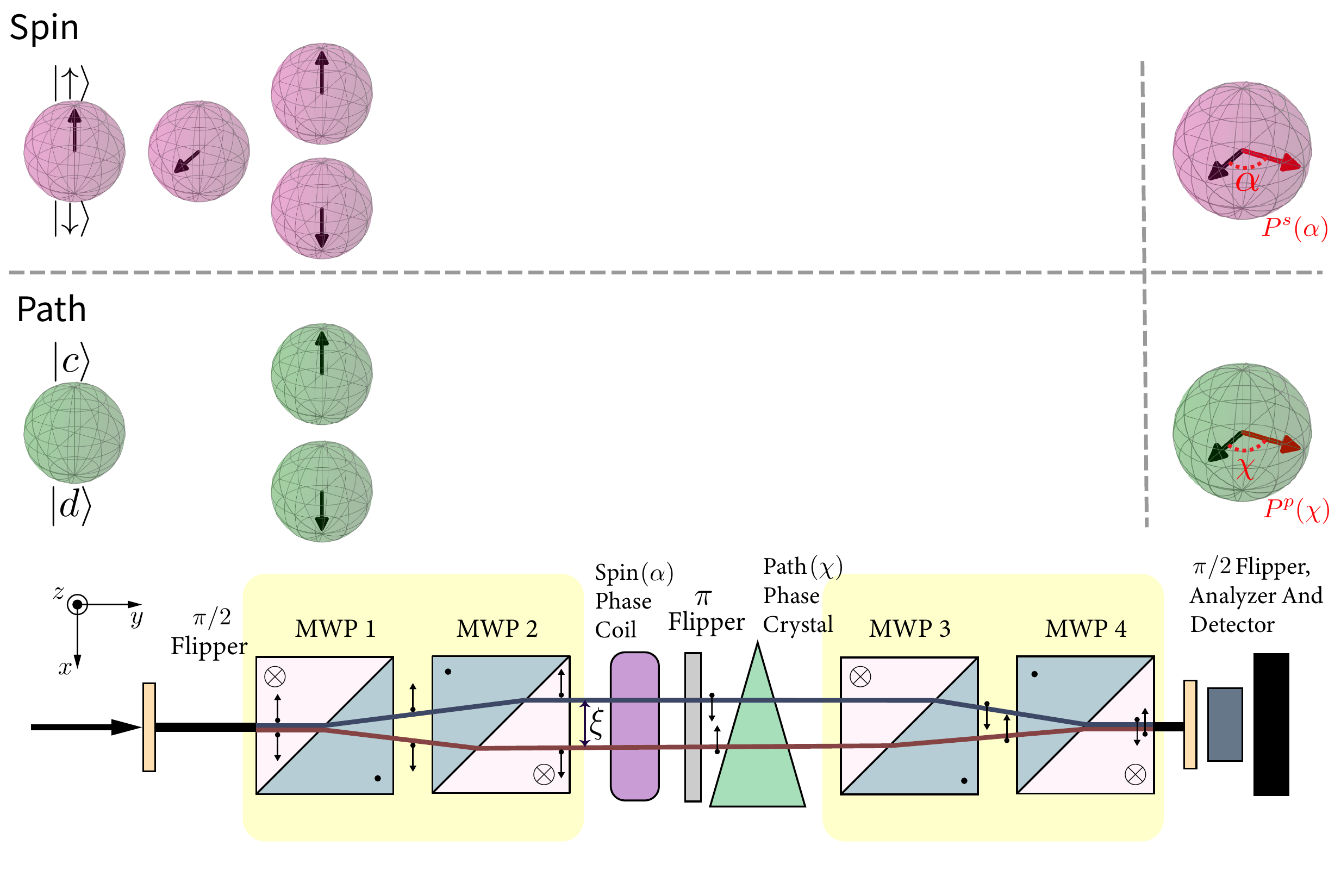}
    \hspace*{1.8cm} \includegraphics[scale=0.35]{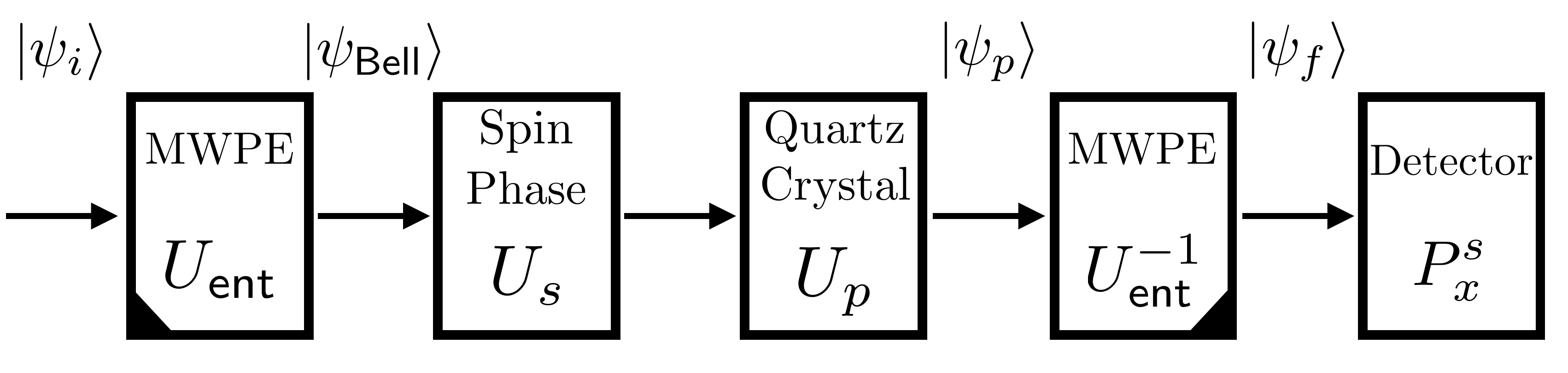}
    \caption{The interferometer consists of two MWPEs, two coherent neutron optical phase generators, and a spin projection measurement. $U_{\sf ent}$ entangles the incident neutron state into spin and path subsystem modes, while $U_{\sf ent}^{-1}$ recombines and disentangles the neutron state. Two phase-shifters $U_s(\alpha)$ and $U_p(\chi)$ acting on spin and path subspaces are applied between the entanglers to generate the phase shifts required for the entanglement witness measurements. A spin analyzer set to pass incident neutrons polarized in the $+x$-direction is followed by a neutron detector to complete the interferometric measurement.}
    \label{Fig: MWP-Interferometer-Flowchart}
\end{figure*}


Next we describe the mode of operation of this entangler when only the spin and path subspaces become entangled. When $\Delta=0$, $E_-=E_0=E_+$, and we can disregard the energy subsystem altogether. The transition amplitudes 
\begin{eqnarray}
    \ket{\up\, 0}&\mapsto& \ket{\up \, 1 },   \label{Eq: transition_up_RF2}\\
    \ket{\down\, 0}&\mapsto& \ket{\down \, 2} \label{Eq: transition_down_RF2}
\end{eqnarray}
are depicted in Fig.~\ref{Fig: RF(2)-Unitary}, for which we propose a Hamiltonian
\begin{equation} \nonumber
    H=\ket{\up \, 1}\bra{\up \, 0} +\ket{\down \, 2}\bra{\down \, 0}+\mbox{h.c.},
\end{equation}
which implies $H^2=\mathds 1-P_{\up 2}-P_{\down 1}$. We follow the same procedure as in the three subsystem case. Expanding the exponential $ \exp[\frac{-i H t}{\hbar}]$ and setting $t=\frac{\pi \hbar}{2}$ one can show that
\begin{equation} \nonumber
    \exp[\frac{-i \pi H}{2}]=P_{\up 2}+P_{\down 1}-i\left(\ket{\up \, 1}\bra{\up \, 0}    + \ket{\down \, 2}\bra{\down \, 0}+\mbox{h.c.}\right).
\end{equation}
Including Larmor precession one can finally derive the desired unitary operator 
\begin{multline}\nonumber
    U_{\sf ent}=e^{ i \varphi} \left(P_{\up 2}-i \ket{\up \, 1}\bra{\up \, 0}-i\ket{\up \, 0}\bra{\up  \, 1}  \right)\\
    +e^{-i \varphi}\left(P_{\down 1}-i \ket{\down \, 2}\bra{\down \, 0}-i\ket{\down \, 0}\bra{\down \, 2}\right).
\end{multline}
To avoid the complex phases in the transition amplitudes discussed in Eq. \eqref{Eq: transition_up_RF2} and \eqref{Eq: transition_down_RF2}, one can again redefine the base states as $-i e^{i\varphi}\ket{1}\rightarrow \ket{1}$ and $-i e^{-i\varphi}\ket{2}\rightarrow \ket{2}$.
\begin{multline}\nonumber
    U_{\sf ent}=\ket{\up \, 1}\bra{\up \, 0}+e^{i \varphi} P_{\up 2}-e^{ i 2\varphi}\ket{\up \, 0}\bra{\up  \, 1}  \\
    +\ket{\down \, 2}\bra{\down \, 0}+e^{-i \varphi}P_{\down 1}-e^{ -i 2\varphi}\ket{\down \, 0}\bra{\down \, 2}.
\end{multline}

For an incident neutron in the state $\frac{\ket{\up}+\ket{\down }}{\sqrt 2}\ot \ket{0}$ with energy $E_0$ this entangler creates a Bell state which in that basis is
\begin{equation} \label{Eq: Bell state prep_RF}
    U_{\sf ent} \frac{\ket{\up}+\ket{\down }}{\sqrt 2}\ot \ket{0}=  \frac{\ket{\up  \, 1 }+\ket{\down \, 2 }}{\sqrt 2}.
\end{equation}

\section{neutron interferometers}\label{Sec: Interferometer}

In this section we construct the mathematical representation of the unitary operations realized by the neutron interferometer used in our measurements by applying sequences of the two entangler operators constructed in Sec. ~\ref{Sec: Entangler}. In combination with coherent neutron optical elements introduced into the interferometer which introduce adjustable phase shifts between the components of the different subsystems, one can realize measurements of the entanglement witnesses described below solely through count rates measured in a neutron detector after a correctly-chosen final state spin projection. 

\subsection{Neutron Interferometer with Magnetic Wollaston Prisms} 
\label{Sec: MWP interferometer}

In this apparatus, entanglement is created using the entangler constructed by a pair of MWPs discussed in Sec. ~\ref{Sec: MWP construction}. 
This apparatus entangles the neutron spin and path subspaces. The interferometer possesses three stages (see Fig. 
\ref{Fig: MWP-Interferometer-Flowchart}). First a Bell state is 
created from an incident unentangled state $\ket{\psi_i}=\frac{\ket{\up }+\ket{\down}}{\sqrt 2}\ot \ket{a}$ by the entangler discussed 
in Eq.~\eqref{Eq: Bell state prep_MWP}  
\begin{equation} \nonumber
    \ket{\psi_{\sf Bell}}=U_{\sf ent} \ket{\psi_i}= e^{i\varphi} \frac{\ket{\up \, c }+ \ket{\down \, d }}{\sqrt 2}.
\end{equation}

To implement the entanglement witness measurements we use a spin phase coil $U_s(\alpha)$, to introduce a relative phase shift between the two spin states, and transmission through the quartz crystal $U_p(\chi)$, to introduce a relative phase between the two path states, 
\begin{eqnarray}
    U_s(\alpha)&=&\ket{\up}\bra{\up}+e^{i \alpha}\ket{\down}\bra{\down},   \nonumber\\
    U_p(\chi)&=&\ket{c}\bra{c}+e^{i \chi}\ket{d}\bra{d} . \nonumber
\end{eqnarray}
The combined effect of these two commuting phase shifters leads to
\begin{equation} \nonumber
\ket{\psi_p}=U_{p} \, U_{s}\st{\sf Bell}
=e^{i\varphi}\frac{\ket{\uparrow c}+e^{i(\alpha+\chi)}\ket{\downarrow d}}{\sqrt 2} .
\end{equation}

The recombination of the path amplitudes of the entangled neutron state is done by an inverse MWP (dis)entangler $U_{\sf ent}^{-1}$ 
\begin{equation} \nonumber
\st{f}=U_{\sf ent}^{-1}\st{p}=\frac{\ket{\uparrow}+e^{i(\alpha+\chi)}\ket{\downarrow}}{\sqrt 2}\otimes\ket{a}.
\end{equation}

Finally, $\ket{\psi_f}$ passes through a $\frac{\pi}{2}$ spin-turner and then enters the polarization analyzer and the detector. 
Each Pauli matrix defined in Eq. \eqref{Eq: Def_Pauli_spin} and \eqref{Eq: Def_Pauli_path} can be decomposed in terms of projectors 
\begin{eqnarray}
    \s^s_{u(\alpha)}&=&P^s(\alpha)-P^s(\alpha+\pi), \label{Eq: Pauli_spin_Decomp}\\
    \s^p_{v(\chi)}&=&P^p(\chi)-P^p(\chi+\pi),\label{Eq: Pauli_path_Decomp}
\end{eqnarray}
defined as  
\begin{eqnarray}
    P^s(\alpha)&=&\ket{+,\alpha}\bra{+,\alpha} \ \mbox{with} \  \ket{+,\alpha}=\frac{\ket{\up}+e^{i \alpha}\ket{\down}}{\sqrt 2} ,   \nonumber \\
    P^p(\chi)&=&\ket{+,\chi}\bra{+,\chi} \ \mbox{with} \  \ket{+,\chi}=\frac{\ket{c}+e^{i \chi}\ket{d}}{\sqrt 2}, \nonumber
\end{eqnarray}
for which $\s^s_{u(\alpha)}\ket{+,\alpha}=(+1)\ket{+,\alpha}$ and $\s^p_{v(\chi)}\ket{+,\chi}=(+1)\ket{+,\chi}$. 
The combined effect of the instruments involved in this last stage realizes a projective measurement $P^s(0)$ on the spin subsystem, which counts neutrons in the state $\frac{\ket{\up}+\ket{\down}}{\sqrt 2}$.  

Let $N(\alpha, \chi)$ be the number of  neutrons detected for phase shifts $\alpha$ and $\chi$, then one can show that
\begin{equation}
    \frac{N(\alpha, \chi)}{N(\alpha, \chi)_{\sf max}}=\frac{|\bra{\psi_f}\ket{+}|^2}{|\bra{+}\ket{+}|^2}=\frac{1}{2}[1+\cos{(\alpha+\chi)}] . \nonumber
\end{equation}
Therefore,
\begin{equation*}
N(\alpha, \chi)\propto 1+\cos(\alpha+\chi).
\end{equation*}
The disentangler $U_{\sf ent}^{-1}$ together with the path phase shift $U_p$ act like a projective measurement $P^p(\chi)$ on the path subsystem,  and the projective measurement $P^s(0)$ together with the spin phase shift $U^s(\alpha)$ act on the spin  like the projective measurement $P^s(\alpha)$. Therefore, 
\begin{equation} \label{Eq: counting statistics_MWP}
    N(\alpha, \chi) \propto E[P^s(\alpha) P^p(\chi)],
\end{equation}
where
\begin{eqnarray*}
    E[P^s(\alpha) P^p(\chi)]&=&\bra{\psi_{\sf Bell}}P^s(\alpha) P^p(\chi)\ket{\psi_{\sf Bell}}\\
    &=&\frac{1}{4}[1+\cos{(\alpha+\chi)}] .
\end{eqnarray*}

\subsection{Neutron Interferometer with RF Flippers}\label{Sec: RF-Flipper interferometer}

\begin{center}
\begin{figure*}[htb!]
    \includegraphics[scale=0.586]{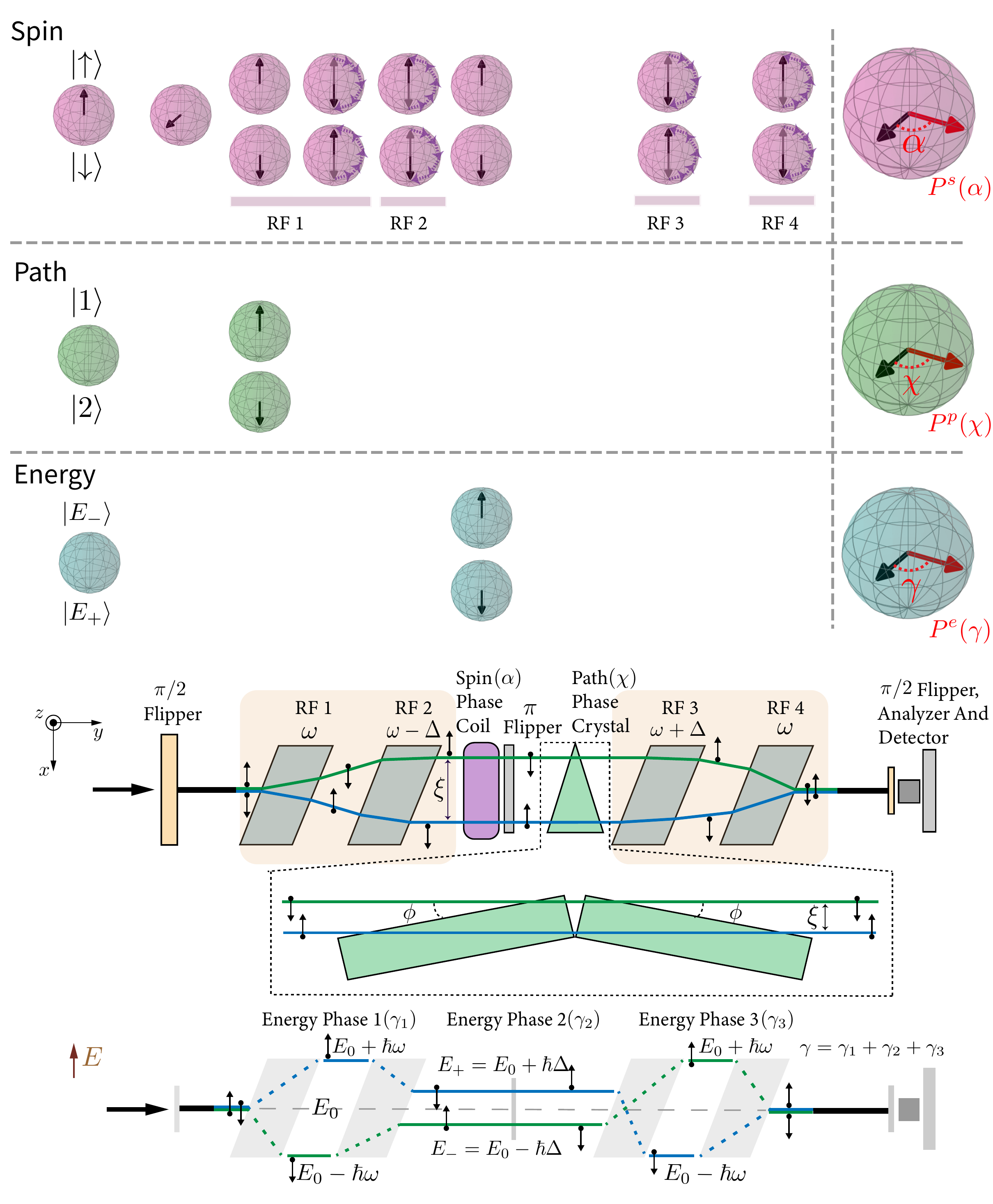}
    \hspace*{2cm}\includegraphics[scale=0.38]{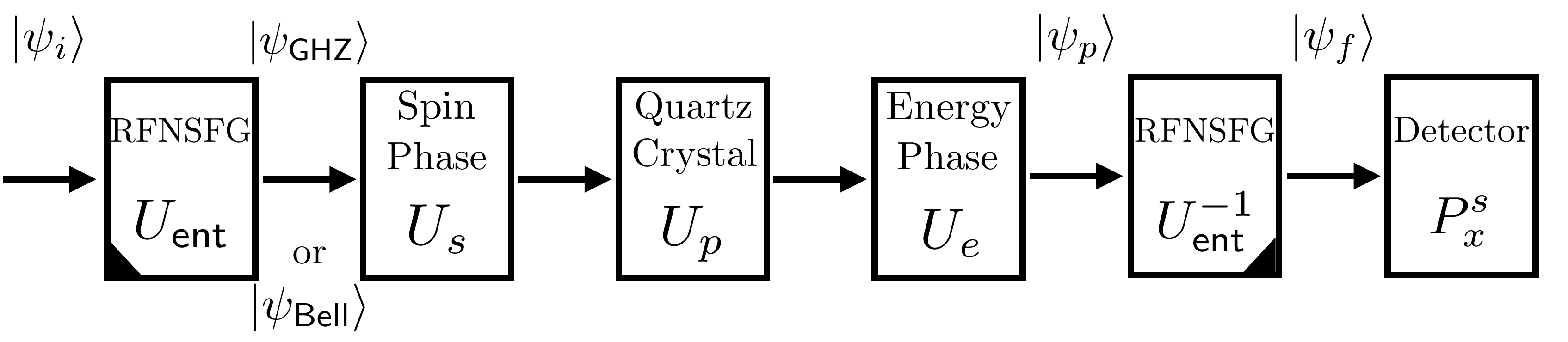}
    \caption{This interferometer consists of two RFNSFG entanglers, three commuting neutron optical phase shifters in the three different distinguished subspaces, and a polarization analyzer and neutron detector. The first RFNSFG, $U_{\sf ent}$, entangles the neutron into spin, path, and energy modes (or, only spin and path modes when $\Delta=0$), while the second RFNSFG, $U_{\sf ent}^{-1}$, recombines and disentangles the neutron state. Three commuting phase shifters $U_s(\alpha)$,  $U_p(\chi)$ and $U_e(\gamma)$ (or, two phase-shifters $U_s(\alpha)$ and  $U_p(\chi)$ for the two subsystem case) are inserted between the flippers to apply phase shifts in the different subsystems. Finally the neutron state is analyzed by the polarization analyzer and counted in the detector. The inset for the path phase crystal shows how this phase can be realized and adjusted using a pair of quartz blocks with an adjustable angle $\phi$. }
    \label{Fig: RF-Interferometer-Flowchart}
\end{figure*}
\end{center}

In this interferometer entanglement is generated using the RFNSFG discussed in Sec. ~\ref{Sec: RF-flipper construction}. This apparatus is very similar to the MWP-based interferometer but it is more flexible as it can entangle either two or three degrees of freedom depending on the mode of operation chosen through the selection of the RF frequencies. We first discuss the three-mode interferometer setup (see Fig. \ref{Fig: RF-Interferometer-Flowchart}). In the first stage a GHZ state is created from an incident unentangled state $\ket{\psi_i}=\frac{\ket{\up }+\ket{\down} }{\sqrt 2}\ot \ket{0 \, E_0}$ 
by the entangler shown in Eq.~\eqref{Eq: GHZ state prep_RF}, 
\begin{equation} \nonumber
\st{\sf GHZ}=U_{\sf ent}\st{i}= \frac{\ket{\up \,  1  \, E_-}+\ket{\down  \,  2  \, E_+}}{\sqrt 2} .
\end{equation}

As for the MWP case we introduce relative phase shifts to perform projective measurements in some desired regions of the subsystems' Bloch spheres.  Three different commuting phase shifters act on the spin, path, and energy modes. Spin and path  phase shifts are generated as above by the spin phase coil and the quartz crystal, respectively. The energy  phase shift is created by a so-called zero-field precession \cite{ggk-1994,sponar-2008} by changing the distance between the entanglers. We denote these three phase shifts as $U_s(\alpha)$, $U_p(\chi)$ and $U_e(\gamma)$ respectively,
\begin{eqnarray}
    U_s(\alpha)&=&\ket{\up}\bra{\up}+e^{i \alpha}\ket{\down}\bra{\down},  \nonumber \\
    U_p(\chi)&=& \ket{1}\bra{1}+e^{i \chi}\ket{2}\bra{2} ,  \nonumber \\
    U_e(\gamma)&=&\ket{E_-}\bra{E_-}+e^{i  \gamma}\ket{E_+}\bra{E_+},  \nonumber
\end{eqnarray}
where we disregard the energy state $\ket{E_0}$ and path state $\ket{0}$ as the neutron does not exist in those states in between the first and last RFNSFs in this apparatus configuration. The neutron state after the phase shifters becomes
\begin{eqnarray}
    \st{p}&=&U_{e}\, U_{p}\, U_{s}\st{\sf GHZ}   \nonumber \\
    &=& \frac{\ket{\uparrow \, 1 \, E_-}+e^{i(\alpha+\chi+\gamma)}\ket{\downarrow \, 2 \, E_+}}{\sqrt 2}.    \nonumber
\end{eqnarray}

These phase-shifts are followed by a recombination of path and energy subspaces of the entangled beam to create an unentangled product state using an inverse RFNSFG
\begin{equation} \nonumber
    \st{f}=U_{\sf ent}^{-1}\st{p} 
    =\frac{\su+e^{i(\alpha+\chi+\gamma)}\sd}{\sqrt 2}\otimes \ket{0 \, E_0}.
\end{equation}


As in the case of the MWP interferometer $\ket{\psi_f}$ passes through the $\frac{\pi}{2}$ spin turner, polarization analyzer, and detector. 
Similar to previous section, we write the Pauli matrices in terms of projectors 
\begin{eqnarray}
    P^s(\alpha)&=&\ket{+,\alpha}\bra{+,\alpha} \ \mbox{with} \  \ket{+,\alpha}=\frac{\ket{\up}+e^{i \alpha}\ket{\down}}{\sqrt 2} ,   \nonumber \\
    P^p(\chi)&=&\ket{+,\chi}\bra{+,\chi} \ \mbox{with} \  \ket{+,\chi}=\frac{\ket{1}+e^{i \chi}\ket{2}}{\sqrt 2} \nonumber,\\
    P^e(\gamma)&=&\ket{+,\gamma}\bra{+,\gamma} \ \mbox{with} \  \ket{+,\gamma}=\frac{\ket{E_{-}}+e^{i \gamma}\ket{E_{+}}}{\sqrt 2}, \nonumber
\end{eqnarray}
with the addition, to Eqs. \eqref{Eq: Pauli_spin_Decomp} and \eqref{Eq: Pauli_path_Decomp}, of $\s^e_{w(\gamma)}=P^e(\gamma)-P^e(\gamma+\pi)$. Those instruments realize the projective measurement $P^s(0)$ on the spin subsystem, which counts neutrons in the state $\frac{\ket{\up}+\ket{\down}}{\sqrt 2}$ incident on the spin turner. 

Let $N(\alpha, \chi, \gamma)$ be the number of neutrons counted with spin, path and energy phase shifts set to $\alpha$, $\chi$ and $\gamma$, then one can show that\begin{equation}
    \frac{N(\alpha, \chi,
    \gamma)}{N(\alpha, \chi,\gamma)_{\sf max}}=\frac{|\bra{\psi_f}\ket{+}|^2}{|\bra{+}\ket{+}|^2}=\frac{1}{2}[1+\cos{(\alpha+\chi+\gamma)}]. \nonumber
\end{equation}
Therefore,
\begin{equation*}
N(\alpha, \chi,\gamma)\propto 1+\cos(\alpha+\chi+\gamma).
\end{equation*}
The disentangler $U_{\sf ent}^{-1}$ together with the path phase shift $U_p$ and energy phase shift $U_e$ act like a combination of two compatible projective measurements $P^p(\chi) $ and $  P^e(\gamma)$ on the path and energy mode, respectively,  and the projective measurement $P^s(0)$ together with the spin phase shift $U^s(\alpha)$ act on the spin like the projective measurement $P^s(\alpha)$. Therefore, 
\begin{equation} 
    N(\alpha, \chi, \gamma) \propto E[P^s(\alpha) P^p(\chi)P^e(\gamma)],
\label{Eq: counting statistics_RF3}
\end{equation}
where
\begin{eqnarray*}
    E[P^s(\alpha) P^p(\chi)P^e(\gamma)]&=&\bra{\psi_{\sf GHZ}}P^s(\alpha) P^p(\chi)P^e(\gamma)\ket{\psi_{\sf GHZ}}\\
    &=&\frac{1}{8}[1+\cos{(\alpha+\chi+\gamma)}] .
\end{eqnarray*}

The two subsystem mode is a limiting case of the three subsystem configuration as discussed in Sec.~\ref{Sec: RF-flipper construction} when $\Delta=0$. In the first stage a Bell state is created from the unentangled incident state $\ket{\psi_i}=\frac{\ket{\up}+\ket{\down}}{\sqrt 2}\otimes\ket{a}$ by the entangler shown in Eq.~\eqref{Eq: Bell state prep_RF}, 
\begin{equation} \nonumber
    \st{\sf Bell}=U_{\sf ent}\st{i}= \frac{\ket{\up \,  1}+\ket{\down  \,  2 }}{\sqrt 2}.
\end{equation}

In this case zero-field precession is absent and the energy phase shifter is the identity $U_e=\mathds 1$. Combining the effects from the spin and path phase shifts one can show that
\begin{equation} \nonumber
    \st{p}=U_{p} \, U_{s}\st{\sf Bell} = \frac{\ket{\up \,  1  }+e^{i(\alpha+\chi)}\ket{\down  \,  2}}{\sqrt 2}.
\end{equation}

In the final stage, the second RFNSF disentangles the state
\begin{equation} \nonumber
    \st{f}=U_{\sf ent}^{-1}\st{p}
    =\frac{\su+e^{i(\alpha+\chi)}\sd}{\sqrt 2}\ot \ket{0} ,
\end{equation}
and $\ket{\psi_f}$ passes through the spin turner, polarization analyzer and detector. The analysis of the neutron count rate $N(\alpha, \chi)$ is 
identical to the one done in Sec. \ref{Sec: MWP interferometer}.

\ignore{
Together they act like a projective measurement $P^s(0)$ to detect and count neutrons in the $\frac{\ket{\up}+\ket{\down}}{\sqrt 2}$ state.  Let $N(\alpha, \chi)$ be the neutron count rate with spin and path phase shifts $\alpha$ and $\chi$, then one can show that
\begin{equation}
    \frac{N(\alpha, \chi)}{N(\alpha, \chi)_{\sf max}}=\frac{|\bra{\psi_f}\ket{+}|^2}{|\bra{+}\ket{+}|^2}=\frac{1}{2}[1+\cos{(\alpha+\chi)}] . \nonumber
\end{equation}
Therefore,
\begin{equation*}
N(\alpha, \chi)\propto 1+\cos(\alpha+\chi),
\end{equation*}
where
\begin{eqnarray*}
    E[P^s(\alpha) P^p(\chi)]&=&\bra{\psi_{\sf Bell}}P^s(\alpha) P^p(\chi)\ket{\psi_{\sf Bell}}\\
    &=&\frac{1}{4}[1+\cos{(\alpha+\chi)}] .
\end{eqnarray*}
Note that $N(\alpha, \chi)$ is proportional to $\|P^s(\alpha) P^p(\chi)|\psi_{\sf Bell} \rangle\|^2$. The reason is that the disentangler $U_{\sf ent}^{-1}$ together with the path phase shift $U_p$ act like a projective measurement $P^p(\chi)$ on the path subsystem,  and the projective measurement $P^s(0)$ together with the spin phase shift $U^s(\alpha)$ act on the spin like the projective measurement $P^s(\alpha)$. Therefore, 
\begin{equation} \label{Eq: counting statistics_RF2}
    N(\alpha, \chi) \propto E[P^s(\alpha) P^p(\chi)].
\end{equation}

}

\section{Statistical Analysis for Two and Three Mode Entanglement}
\label{Sec: Statistical Analysis}

In this section we express the CHSH and Mermin entanglement witnesses defined in Sec.~\ref{Sec: Background} in terms of the interferometer count rates defined in Sec. \ref{Sec: Interferometer} for special choices of the phase shifts. These interferometers cannot directly measure the observables in  Eq.~\eqref{Eq: CHSH Witness} and \eqref{Eq: Mermin-Witness} as the only available data comes from the polarization analysis in the spin subsystem. However one can construct the entanglement witnesses of interest by conducting measurements with different settings of the phase shifters. 

Consider a particular context $\{\sigma^s_{u(\alpha)}, \sigma^p_{v(\chi)}\}$ in a CHSH witness $S$ from the arrangements described in Sec.~\ref{Sec: MWP interferometer} and \ref{Sec: RF-Flipper interferometer}. By decomposing each Pauli matrix into two projectors as in Eq.\eqref{Eq: Pauli_spin_Decomp} and Eq.\eqref{Eq: Pauli_path_Decomp}, one can derive the expectation value of the context $E(\alpha, \chi)=\bra{\psi_{\sf Bell}}\sigma^s_{u(\alpha)} \sigma^p_{v(\chi)}\ket{\psi_{\sf Bell}}$ as
\begin{eqnarray}\nonumber
    E(\alpha, \chi)&=&\frac{\sum_{\mu_s,\mu_p}(-1)^{\mu_s+\mu_p}E\left[P^s(\alpha+\mu_s\pi) P^p(\chi+\mu_p \pi)\right]}{\sum_{\mu_s,\mu_p} E\left[P^s(\alpha+\mu_s \pi)P^p(\chi+\mu_p \pi)\right]}\\
    &=&\frac{\sum_{\mu_s,\mu_p}(-1)^{\mu_s+\mu_p} N(\alpha+\mu_s \pi, \chi+\mu_p \pi)}{\sum_{\mu_s,\mu_p} N(\alpha+\mu_s \pi, \chi+\mu_p \pi)},   \nonumber
\end{eqnarray}
where we use Eq.~\eqref{Eq: counting statistics_MWP}. To determine the expectation value $E(\alpha, \chi)$ one needs measurements with four different phase shift settings $\{N(\alpha+\mu_s \pi, \chi+\mu_p \pi)\}$ with $\mu_s, \mu_p =0,1$. We expect the maximum violation of the CHSH inequality in Eq.~\eqref{Eq: CHSH Witness} when $\alpha_1 + \chi_1 = -\frac{\pi}{4}$ and  $\alpha_2 - \alpha_1 = \chi_2 - \chi_1 = \frac{\pi}{2}$.

We can also determine the expectation values of the relevant contexts involved in 
the Mermin witness $M$ defined in Eq.~\eqref{Eq: Mermin-Witness}, $\{\sigma^s_{u(\alpha)}, \sigma^p_{v(\chi)}, \sigma^e_{w(\gamma)}\}$ with $\alpha, \chi, \gamma=0,\frac{\pi}{2}$. We discussed the three subsystems case in Sec.~\ref{Sec: MWP interferometer}. By decomposing the Pauli matrices into projectors like in the two subsystems case we get 
\begin{eqnarray}\nonumber
    E(\alpha, \chi, \gamma)&=&\frac{\sum_{\mu_s, \mu_p, \mu_e } (-1)^{\mu_s+\mu_p+\mu_e} E_{\mu_s\mu_p \mu_e}}{\sum_{\mu_s, \mu_p, \mu_e } E_{\mu_s\mu_p \mu_e}}\\
    &=&\frac{\sum_{\mu_s, \mu_p, \mu_e } (-1)^{\mu_s+\mu_p+\mu_e} N_{\mu_s\mu_p \mu_e}}{\sum_{\mu_s, \mu_p, \mu_e } N_{\mu_s\mu_p \mu_e}}, \nonumber
\end{eqnarray}
where $E_{\mu_s \mu_p \mu_e}=E[P^s(\alpha+\mu_s \pi) P^p( \chi+\mu_p \pi) P^e( \gamma+\mu_e \pi)]$ and $N_{\mu_s \mu_p \mu_e}=N(\alpha+\mu_s \pi, \chi+\mu_p \pi, \gamma+\mu_e \pi)$ after using  Eq.~\eqref{Eq: counting statistics_RF3}. To determine the expectation value of that context one needs eight measurements with different phase shift settings $\{N(\alpha+\mu_s \pi, \chi+\mu_p \pi, \gamma+\mu_e \pi)\}$ with $\mu_s, \mu_p, \mu_e =0,1$. 

\section{Generalization to Multiple-Mode Entanglement}
\label{Generalization}

Here we generalize the results derived above for a neutron interferometer which possesses $n$ entangled distinguishable subsystems and show how one can determine the expectation value of a context using counting statistics from the detectors of that interferometer assuming that the final stage of the interferometer consists of the same spin projection and neutron detection combination as described above. This calculation might be useful for future entangled neutron state measurements if one can generate single-particle neutron states with OAM whose amplitudes can be treated to a good approximation as finite-dimensional subsystems. In this case it might be possible to form single-particle entangled neutron states in four distinct properties: spin, path, energy, and OAM.     

We express the full Hilbert space in terms of the tensor product decomposition
\begin{equation} \nonumber
    \cH=\bigotimes_{l=0}^{n-1} \cH_l,
\end{equation}
where $\cH_l$ is the two dimensional space for the $l$-th subsystem and $l=0$ labels the spin subsystem. Choose a basis $\{\ket{\Up}, \ket{\Down}\}$ for $\cH_l$. Define the context to be measured as $\{\s_{u(\phi_0)}^{0}, \ldots, \s_{u(\phi_{n-1})}^{n-1}\}$, where 
\begin{equation}\nonumber
    \s_{u(\phi_l)}^{l}=\cos{\phi_l}\, \s^{l}_x+\sin{\phi_l}\, \s^l_y \ \ , \  l=0,1,\cdots, n-1 , 
\end{equation}
acts on the $l$-th subsystem. 
Now, one can decompose $\s_{u(\phi_l)}^{l}$ into projectors as $\s_{u(\phi_l)}^{l}=P^l(\phi_l)-P^l(\phi_l+\pi)$, where 
\begin{equation}    \nonumber
    P^l(\phi_l)=\ket{+,\phi_l}\bra{+,\phi_l}  \ \mbox{with} \ \ket{+,\phi_l}=\frac{\ket{\Up}+e^{i \phi_l}\ket{\Down}}{\sqrt 2}.
\end{equation}
Let the entangler of the interferometer prepare a maximally entangled state $\ket{\psi_{\sf E}}=\frac{\ket{\Up \Up \ldots \Up}+\ket{\Down \Down \ldots \Down}}{\sqrt{2}}$. 
\ignore{For such a state one can show that
\begin{equation} \nonumber
    E\left[\prod_l \s_{w_l}^l\right]=\cos \left(\sum_l\phi_l\right).
\end{equation}}
Let the $l$-th phase shifter $U_l(\phi_l)$ introduce a relative phase $e^{i\phi_l}$ between $\ket{\Up}$ and $\ket{\Down}$,
\begin{equation}\nonumber
    U_l=\ket{\Up}\bra{\Up}+e^{i\phi_l} \ket{\Down}\bra{\Down}.   
\end{equation}
Combining the effects of all the phase shifters one can show that
\begin{equation}    \nonumber
    \ket{\psi_p}=\prod_l U_l \ket{\psi_{\sf E}}=\frac{\ket{\Up \Up \ldots \Up}+e^{i\sum_l\phi_l}\ket{\Down \Down \ldots \Down}}{\sqrt{2}}.
\end{equation}
In the next stage the state is disentangled by $U_{\sf det}$. Although the full mathematical description of this disentangler depends on apparatus details, the relevant action on the state is 
\begin{eqnarray}
   U_{\sf det} \, \ket{\Up \Up \ldots \Up} &=& \ket{\Up}\otimes \ket{\sf unknown} , \nonumber \\
   U_{\sf det} \,  \ket{\Down \Down \ldots \Down} &=& \ket{\Down} \ot \ket{\sf unknown} ,  \nonumber
\end{eqnarray}
where $\ket{\sf unknown}$ is some unknown state. 
We get the final state
\begin{equation}    \nonumber
    \ket{\psi_f}=U_{\sf det} \ket{\psi_p}= \frac{\ket{\Up}+e^{i\sum_l\phi_l}\ket{\Down}}{\sqrt 2}\ot \ket{\sf unknown},
\end{equation}
which passes through the spin-turner and polarization analyzer. Together they realize the projective measurement $P^0(0)$ which detect and count neutrons in the spin state $\frac{\ket{\Up}+\ket{\Down}}{\sqrt 2}$. Let $N(\{\phi_l\})$ be the number of neutrons detected in the detector with phase shifts set to the angles $\{\phi_l\}$. Then 
\begin{equation}
   \frac{ N(\{\phi_l\})}{N(\{\phi_l\})_{\sf max}}=\frac{|\bra{\psi_f}\ket{+}|^2}{|\bra{+}\ket{+}|^2}=\frac{1}{2}[1 +\cos (\sum_l \phi_l)] . \nonumber
\end{equation}
Therefore,
\begin{equation*}
  N(\{\phi_l\})  \propto 1 +\cos (\sum_l \phi_l) .
\end{equation*}
Note that $N(\{\phi_l\})$ is proportional to 
\begin{eqnarray*}
    E\left[\prod_l P^l(\phi_l)\right]&=&\bra{\psi_{\sf E}} \prod_l P^l(\phi_l) \ket{\psi_{\sf E}}\\
    &=&\frac{1}{2^n}[1+\cos (\sum_l \phi_l)] ,
\end{eqnarray*}

The reason is that the disentangler $U_{\sf det}$ together with the  phase shift $\prod_l U_l$ acts like a combination of $n-1$ compatible projective measurements $\{ P^l(\phi_l)\}$, and the projective measurement $P^0(0)$ together with the spin phase shift $U^0(\phi_0)$ act on the spin like the projective measurement $P^0(\phi_0)$.

Using the above result one can derive the expectation value of the context as
\begin{eqnarray}
    E\left[\prod_l \s^l_{u(\phi_l)}\right]&=&\frac{\sum_{\{\mu_l\}}(-1)^{\sum_l \mu_l}E_{\{\mu_l\}}}{\sum_{\{\mu_l\}}E_{\{\mu_l\}}}  \nonumber\\
    &=&\frac{\sum_{\{\mu_l\}}(-1)^{\sum_l \mu_l}N_{\{\mu_l\}}}{\sum_{\{\mu_l\}}N_{\{\mu_l\}}},     \nonumber
\end{eqnarray}
where $E_{\{\mu_l\}}=E[\prod_l P^l(\{\phi_l+\mu_l\pi\})]$ and $N_{\{\mu_l\}}=N\left(\{\phi_l+\mu_l\pi\}\right)$. To determine the expectation value of the context one needs $2^n$ measurements with different phase shifter settings $\{N(\{\phi_l+\mu_l \pi\})\}$ with $\mu_l=0,1$. 

\section{Experimental Results for Two and Three Entangled Subsystems}
\label{Experimental Results}

The construction presented in the previous sections forms the theoretical underpinnings for a quantitative analysis of the recent entanglement witness measurements performed on the Larmor neutron spin-echo instrument at the ISIS Neutron and Muon Source Center in the UK \cite{iu-2019}.  We refer the reader to that paper for details on the apparatus and measurement procedure. The measured values of the CHSH, $S$, and Mermin, $M$, contextual witnesses are listed in Table 
\ref{TableExp}.
\begin{table} [htb]
\begin{center}
\begin{tabular}{ |C{1.7cm}||C{2cm}|C{1.5cm}|C{1.5cm}| }
 \hline
 Contextual Witness & Measured Value & Classical Bound & Quantum Bound\\ 
 \hline
 \multirow{3}{*}{$S$} & $2.16 \pm 0.01 \mbox{(stat)} \pm 0.02 \mbox{(sys)}$ & \multirow{3}{*}{$1.56$} & \multirow{3}{*}{$2.21$}\\ 
 \hline
 \multirow{3}{*}{$M$} & $3.052 \pm 0.007 \mbox{(stat)} \pm 0.017 \mbox{(sys)}$ & \multirow{3}{*}{$1.56$} & \multirow{3}{*}{$3.12$} \\ 
 \hline
\end{tabular}
\end{center}
\caption{Experimental results for the two and three mode-entangled single-neutron interferometer \cite{iu-2019}. The classical bound (see text) is obtained from $2 \times 0.78 =1.56$, while the quantum bounds are 
$2\sqrt{2}\times 0.78=2.21$ and $4 \times 0.78=3.12$.}
\label{TableExp}
\end{table}

The values presented above for the measured entanglement witnesses use for their normalization the measured product, ${\sf Pol} \times {\sf A}=0.78$, of the incident neutron beam polarization $\sf Pol$ of the polarizer, and the analyzing power $\sf A$ of the polarization analyzer. There are several subtle aspects of the neutron beam instrumentation which are subsumed into this product. 
Since individual neutrons come from a distribution of momenta and trajectories, each experiences a slightly different Hamiltonian evolution as it moves through the instrument and individual neutrons therefore experience different final phase differences between their up and down states. To manipulate the neutrons’ spin during the experiment we have to engineer magnetic fields with a particular geometry to change the neutron state (or not) and these do not work with perfect efficiency. The overall result of these instrumental effects is that we lose contrast in our interferograms. We express this loss of contrast by a single number – the polarization product ${\sf Pol} \times {\sf A}$ – which multiplies the result that we would get if the apparatus were ideal to give our actual result. 

We do not have enough information on all of the imperfections of our apparatus to be able to isolate and quantify the relative contributions to the observed entanglement witnesses from decoherence, dephasing, and statistical averaging effects. Such a more detailed analysis and investigation would be required to quantify with higher precision the degree of deviation of our measured entanglement witnesses from the classical and quantum bounds. Given the relative simplicity of the neutron interactions with the matter and external fields of the apparatus, however, nothing would preclude in principle such a more detailed (and complex) characterization. Interactions of neutrons are typically weak enough that one can apply either perturbative or coherent optical analyses to the neutron-matter interactions with the apparatus components based on the well-measured neutron scattering amplitudes from atoms in materials.     

Even without such a more detailed analysis, it is clear that our measurement strongly violates the classical bound and is quite close to the expected quantum bound. We conclude that our experiment verifies quantum contextuality in both the double and triple entangled cases and that the single neutron states are entangled. 

\section{Discussion and Outlook}\label{Discussion and Outlook}


The main intellectual motivation for this work is to develop a qualitatively new type of neutron scattering modality which can identify entangled degrees of freedom in matter, without prior knowledge of the responsible many-body interactions that in practice are usually unknown. Although various authors \cite{bls-2000,kl-2002,cle-2003,wmgkhb-2004,hw-2007} recognize the need for the development of a theory for entangled-particle scattering from entangled matter, to our knowledge only partial steps along various lines have been taken so far \cite{squires-book}. It should be obvious that such a theory must exist and cannot be in danger of violating any of the fundamental assumptions of quantum mechanics. 
Particles in many-body systems generally become entangled upon scattering, and the system which has undergone many internal scattering events among its constituent parts is represented by an entangled state. A process in which one entangled particle comes into and goes out of such a system, which we refer to as a ``scattering" experiment, is clearly just a special case of this same type of physical process. One of the ways in which the textbook scattering theory formalism breaks down in the case of entangled particle scattering is the so-called ``cluster decomposition" assumption, which says that the results of macroscopically-separable experiments produce uncorrelated results \cite{mutze-1978}. This condition is obviously violated by entanglement-sensitive witnesses such as those involved in CHSH, GHZ \cite{ghz-1989}, and Mermin type inequalities. From a scattering theory perspective, one can view the existing measurements of those inequalities  using neutron interferometry as entangled neutron ``scattering" from an unentangled system in the forward scattering limit in which the internal state of the matter and external fields, used to manipulate the subsystems, is unchanged. 

The information encoded in the entanglement of the state of the system is generally not directly accessible to the type of probes  developed in the twentieth century. These probes were conceived to investigate the properties of quasiparticles of energy $\hbar \omega$ and momentum $\vec{q}$, according to the Landau paradigm of elementary excitations, without additional theoretical knowledge of the excitations and interactions in the system. The quantitative interpretation of scattering probes of condensed matter systems using the linear-response-based van Hove theory~\cite{VanHove1954} factorizes the double differential scattering cross section ${d^{2}\sigma} \over {d\Omega dE}$ of the process into a product of the scattering amplitudes from individual objects in the system, and the static $S(\vec{q})$ or dynamic $S(\vec{q}, \omega)$ structure factors. In turn $S(\vec{q})$ and $S(\vec{q}, \omega)$ can be expressed in terms of expectation values of various types of space and time correlation functions of those system properties which couple linearly to the probe. This treatment assumes that the probe possesses no entanglement and that one defines the initial and final scattering states of the probe in terms of the same types of unentangled single-particle wave packets assumed in traditional nonrelativistic scattering theory~\cite{Newton, Taylor}. The correlations revealed in this type of scattering measurement include both ``classical" many-body correlations as well as correlations which could be due to quantum entanglement in the interacting system, but there is no way using the traditional single particle scattering measurements of $S(\vec{q})$ and $S(\vec{q}, \omega)$ to uniquely identify the component from the observed correlations which quantum entanglement might be responsible for.  


The question is whether or not one can formulate a theory in a sufficiently general way, to deliver physically interesting information about a system whose specific Hamiltonian is not known in advance, so that entangled particle scattering can be used as a multipurpose scientific tool.  We are presently engaged in this theoretical construction for the case of neutron scattering.
This is no accident. Neutrons are an excellent choice for such a theoretical and experimental development. The zero electric charge, small magnetic moment, and very small electric polarizability of the neutron make it highly insensitive to many sources of environmental decoherence which can threaten to ruin interferometric measurements. The entangled states of neutrons which we have created and characterized in this work are highly robust as shown by the near-saturation of the entanglement witness quantum bounds, despite the transmission of the neutrons through macroscopic amounts of matter in the apparatus. The range of energies and momenta used in neutron scattering measurements for condensed matter and materials research lies well below the thresholds for the ionization of matter, thereby allowing coherent interactions of the neutron with matter to play a dominant role.

From this perspective, one can view the formalism developed in this paper as a zeroth-order version of this eventual entangled scattering theory 
for forward elastic interactions with unentangled systems. Indeed the procedures by which the various phase shifts of the different entangled modes 
are implemented 
all involve interactions with either macroscopic classical fields in coherent states of the electromagnetic field or with the macroscopic neutron optical potential of matter. In both of these cases any entanglement which might be generated in the interaction of the neutrons with these media are coupled to macroscopic collective coordinates of the apparatus. These collective coordinates suffer such rapid decoherence from interactions with the environment that for all practical purposes they can be treated as semi-classical external fields acting on the neutron's quantum degrees of freedom. As a result there is no need to treat explicitly their internal degrees of freedom and we can model them simply as ``entanglers" of the neutron degrees of freedom. Furthermore there is no intrinsic pre-existing entanglement present in these media which couples to the neutron. Coherent states of the electromagnetic field are 
semi-classical, and the averaging procedure used in the multiple scattering theory which forms the foundation of the concept of the neutron optical potential~\cite{Sears} implicitly erases the effects of any unknown entanglement dynamics which might be present in the medium. Therefore, in the limit in which the neutron interactions with the entanglement-generating devices are coherent there is in principle no danger that the entanglers themselves will pollute the interpretation of a future entangled neutron scattering theory from entangled systems.                      


The interferometric measurements with entangled neutrons generated by the entanglers described and modeled in this paper enable one to probe condensed matter phenomena on small spatial scales. In this paper, we described single-particle multimode-entanglement dynamics using two entanglement-generating neutron devices: the MWP and the RFNSF. Using a single neutron-quantum optics analysis, we derived the theoretical expressions needed to quantify contextuality to deal with entangled neutron beams generated by these devices. We constructed the CHSH witness for the doubly-entangled states in spin and position and the Mermin witness for the triply-entangled states in spin, position, and energy. We also showed how the Mermin witness reduces to the CHSH witness when the energy shift $ \hbar \Delta$ vanishes. 
We controlled the entanglement length $\xi$ at micron length scales and the wavelength splitting to $0.01$nm. Our ability to control and vary the entanglement lengths and neutron wavelength splitting on these scales, which correspond to the scales of many different types of dynamical phenomena in condensed matter, can provide us with a qualitatively new probe of correlated materials such as unconventional superconductors, frustrated magnets hosting quantum spin liquid phases and exotic chiral orders. We can certainly explore the fundamental properties of the quantum world in new regimes.

At the experimental level, the immediate continuation of our program to develop entangled neutron probes requires a more detailed understanding of what happens to the interference contrast as one ``turns off" the entanglement in the energy or path subsystems. The detailed behavior of the interference contrast in this regime cannot be modeled within the discrete approximation used in this paper. As the modes' variables overlap the dynamics becomes sensitive to the form of the quantum amplitudes, which are really continuous functions of the dynamical variables and depend on the longitudinal and transverse coherence functions of the neutrons in the beam. 
Experiments are needed to map out the dynamical range over which the finite discrete mode approximations to the dynamics that we have used for the models presented in this paper are valid. An even more stringent test is the implementation of a quantum self-testing protocol to determine a lower bound in the fidelity of the supposedly multimode-entangled state one wants to generate. Although self-testing protocols using CHSH and Mermin type inequalities for spacelike separated entangled particles are known \cite{Scaranibook}, that is not the case for the timelike entangled single-particle states of interest in our work.

We would like next to indicate other possible scientific applications of our entangled neutron states. OAM beams, which possess a nonzero OAM about the beam axis and are therefore not the traditional plane wave states described in the scattering theory textbooks, have now been created for photons, electrons, and neutrons~\cite{Clark2015, Sarenac2019}. 
We envision that in the future it may be possible to create single-particle entangled neutron states in four different quantum mechanical variables: spin, spatial position, energy, and OAM. Such a multiply-entangled state of a single neutron has to our knowledge never been created or investigated experimentally. A recent calculation shows that one could produce a single-particle neutron state entangled in spin and OAM using the electromagnetic spin-orbit scattering of neutrons from atoms~\cite{Afanasev2019}.

We also note that entangled neutron interferometry of the type described in this paper can be used to pose and answer new questions regarding the influence of inertial and gravitational effects on entangled particles. One example is the famous Sagnac effect, which consists of an extra phase shift upon recombination of the two paths taken by a particle traversing an interferometer coming from a rotation of the interferometer apparatus. The Sagnac phase shift has been measured using photons~\cite{Sagnac1913, Michelson1925}, neutrons~\cite{Werner1979, Staudenmann1980}, atoms~\cite{Riehle1991}, and other particles and excitations. The usual derivation of the Sagnac effect assumes that the particle spin defines the inertial reference frame, leading to the so-called Fermi-Walker transport of spin, which assumes that the spin of the particle acts effectively like a gyroscope. This definition suffices in the semiclassical limit in which one can integrate the phase shifts over a well-defined loop in space in an unentangled spin state for which one can define the gyroscope direction at any point on the particle trajectory from the location of the spin state on the Bloch sphere. In the case of the spin and position entangled neutron beam in the interferometers discussed in this paper, however, the initial and final vertically polarized neutron spin states are spatially separated into an entangled state of positive and negative helicity neutron spin states whose spin projection along the loop in space around the interferometer trajectory is longitudinal. In this case it is unclear what spin direction should define the frame from which one constructs the Fermi-Walker transport used in the Sagnac effect derivation. We have been unable to find any measurement of the Sagnac phase shift performed using spin-entangled beams using any  type of matter-wave interferometer. Entangled neutron interferometers of the type described in this paper could be employed to search for the Sagnac effect using a spin-entangled beam of massive particles. Such work would complement similar investigations in progress involving entangled photons on entanglement effects in noninertial frames~\cite{Fink2017, Fink2019, Restuccia2019, Toros2019}.

\begin{acknowledgements}

We thank Prof. Y. Hasegawa for useful discussions and for providing a figure of the interferometer used in the first three-mode single-particle neutron quantum contextuality tests. Experiments at the ISIS Neutron and Muon Source were supported by a beamtime allocation RB182019220 from the Science and Technology Facilities Council. W.M.S acknowledges NSF PHY-1614545, NSF PHY-1913789, and the IU Center for Spacetime Symmetries. A number of the authors acknowledge support from the US Department of Commerce through cooperative agreement number 70NANB15H259. The IU Quantum Science and Engineering Center is supported by the Office of the IU Bloomington Vice Provost for Research through its Emerging Areas of Research program. The work described in this paper arose from the development of magnetic Wollaston prisms funded by the US Department of Energy through its STTR program (grant number DE-SC0009584). 

\end{acknowledgements}
\bibliographystyle{apsrev4-1.bst}
\bibliography{library.bib}

\clearpage
\newpage
\appendix

\section{Triple Neutron Entanglement with a Single RF Flipper}
\label{Appendix: Hasegawa}

\begin{center}
\begin{figure}[h]
    \includegraphics[scale=0.45]{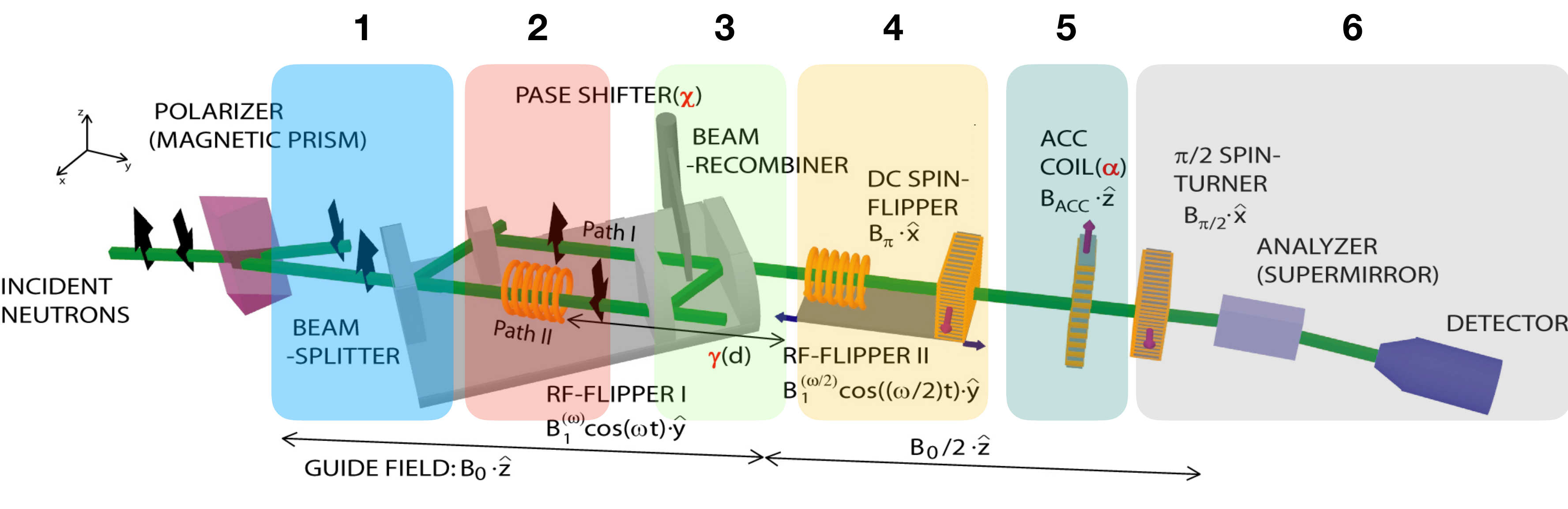}
    \hspace*{1.0cm} \includegraphics[scale=0.40]{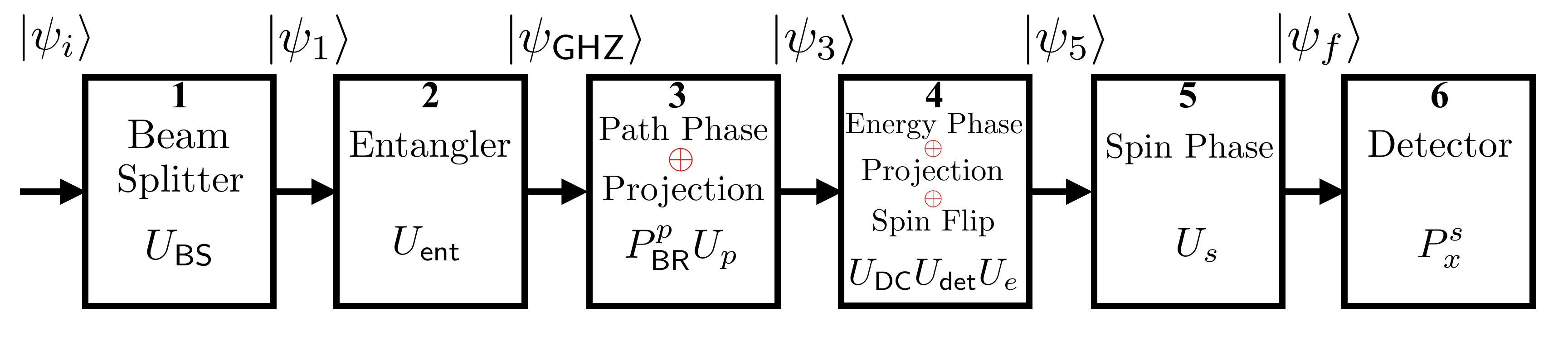}
    \caption{Top image: The experimental apparatus used in Ref.~\cite{hase-2010}. Bottom image: flow chart for the three-subsystem entangled states created and measured in this experiment using the notation of this paper.}
    \label{Set-up: Hasegawa}
\end{figure}
\end{center}

Here we apply the formalism developed in this work to an earlier neutron entanglement experiment performed using a perfect crystal neutron interferometer by Hasegawa {\it et. al.} (see Fig. \ref{Set-up: Hasegawa}). A triply-entangled neutron state was prepared and analyzed using this device in 2010~\cite{hase-2010}. We discuss the construction and working principle of this perfect crystal interferometer and describe how to calculate the contexts for the Mermin witness within the formalism developed in the main text. A very similar analysis could also be applied to several other perfect crystal neutron interferometry experiments.


\onecolumngrid

\twocolumngrid

\begin{figure}[htb]
\includegraphics[scale=0.35]{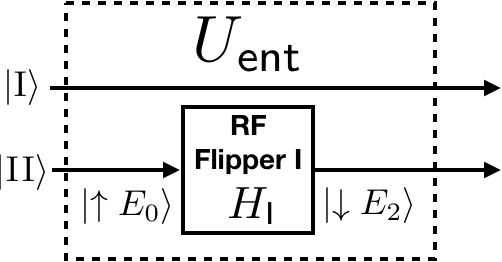}
 \caption{Formation of the entangler.}
 \label{hase-entangler}
\end{figure}

A neutron with spin up and with energy $E_0$ is created from an initially unpolarized monochromatic neutron beam by a polarizer, and is incident upon a 50:50 perfect crystal beamsplitter. Define the incident path as $\mbox{\sc II}$ and the reflected path as $\mbox{\sc I}$, then
\begin{equation} \label{Eq: BS_Hasegawa}
    U_{\sf BS}=\frac{\ket{\I}\bra{\I}+\ket{\I}\bra{\II}+\ket{\II}\bra{\I}-\ket{\II}\bra{\II}}{\sqrt{2}}
\end{equation}
creates the state for subsequent entanglement operations
\begin{equation} \nonumber
    \st{1}=U_{\sf BS}\st{i}=\ket{\up}\ot \frac{\ket{\I}-\ket{\II}}{\sqrt 2}\ot\ket{E_0} .
\end{equation}

The entangled state is generated by a RFNSF operating on path II with the frequency $\omega$. As discussed in Sec.~\ref{Sec: RF-flipper construction}, a RFNSF flips the spin of an incoming neutron  by exchanging a photon of the operating frequency of the RF field. This process is described by the Hamiltonian
\begin{equation} \nonumber
    H_{\sf I}=\ket{\downarrow\, \II \, E_2}\bra{\uparrow \, \II \, E_0}+\mbox{h.c.},
\end{equation}
where $E_2=E_0-\hbar \omega$. By expanding the propagator $U_{\sf ent}=\exp[\frac{-i H_{\sf I} t}{\hbar}]$ into series and setting $t=\frac{\pi \hbar }{2}$ one can show that
\begin{equation} \nonumber
    U_{\sf ent}=\mathds 1-i H_{\sf I} -H_{\sf I}^2,
\end{equation}
where we used $H_{\sf I}^3=H_{\sf I}$. This propagator performs the transition $\ket{\uparrow \, \II \, E_0}\mapsto\ket{\downarrow \, \II \,  E_2}$. Thus the entangler generates an entangled GHZ state
\begin{equation} \nonumber
\st{\sf GHZ}=U_{\sf ent}\ket{\psi_1}=\frac{\ket{\up \, \I \,  E_0} +i\ket{\down \, \II \,  E_2}}{\sqrt 2}.
\end{equation}

Next a slab of matter imposes a neutron optical potential represented by the operator 
\begin{equation}    \nonumber
    U_{p}(\chi)=\ket{\I}\bra{{\mbox{\sc I}}}+e^{i\chi}\ket{\II}\bra{{\mbox{\sc II}}},
\end{equation} 
which generates the state
\begin{equation}    \nonumber
    \st{2}=U_p(\chi) \st{\sf GHZ}=\frac{\ket{\up \,  {\mbox{\sc I}} \,  E_0}+e^{i(\chi+\frac{\pi}{2})}\ket{\down \, {\mbox{\sc II}}\, E_2}}{\sqrt{2}}.
\end{equation} 
Neutron amplitudes in paths I and II diffract from the second blade of the interferometer and then impact on the third blade of the interferometer. The third blade of the interferometer recombines the beam. On the path $\I$ component of the final state one has an entangled state in spin and energy. This projection operation can be defined as
\begin{equation}    \nonumber
    P^p_{\sf BR}= \cN \ket{\I}\bra{\I}U_{\sf BS}^{\dagger},
\end{equation}
where $\cN$ is a normalization constant. This projection operation leads to the state
\begin{equation}    \nonumber
    \st{3}=P^p_{\sf BR}\st{2}=\frac{\ket{\up \, \I \, E_0}+e^{i(\chi+\frac{\pi}{2})}\ket{\down \, \I \, E_2}}{\sqrt 2}.
\end{equation}
\begin{figure}[htb]
    \includegraphics[scale=0.35]{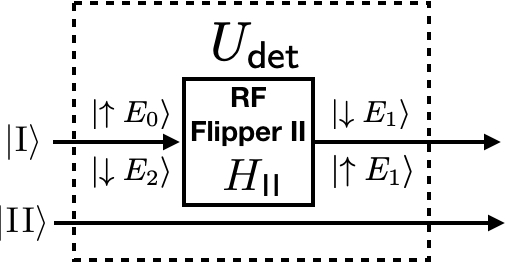}
    \caption{Formation of the disentangler.}
    \label{hase-disentangler}
\end{figure}

In the next stage we perform a projective measurement on the energy mode. With a RFNSF on path $\I$  energy states are recombined into a new energy $E_1=E_0-\frac{\hbar \omega}{2}$ by the transitions $\ket{\uparrow\, \I \, E_0}\rightarrow\ket{\downarrow\, \I \, E_1}$ and
$\ket{\downarrow \, \I \, E_2}\rightarrow \ket{\uparrow \, \I \, E_1}$ using a RFNSF operating with frequency $\frac{\omega}{2}$. The relevant Hamiltonian is 
\begin{equation} \nonumber
    H_{\sf II}=\ket{\downarrow}\bra{\uparrow}\ot\ket{\I}\bra{\I}\ot T+\mbox{h.c.}
\end{equation}
where $T=\E{1}\bra{E_0}+\E{2}\bra{E_1}$. Expand $U_{\sf det}=\exp[\frac{-i H_{\sf II} t}{\hbar}]$ and set $t=\frac{\pi \hbar }{2}$ to obtain
\begin{equation} \nonumber
    U_{\sf det}=\mathds 1-i H_{\sf II} -H_{\sf II}^2,
\end{equation}
where we used $H_{\sf II}^3=H_{\sf II}$. This energy recombination is accompanied by a controllable phase shift in the energy subsystem from zero field precession
\begin{equation}\nonumber
    U_e(\chi)=\ket{E_0}\bra{E_0}+e^{i \gamma} \ket{E_2}\bra{E_2}.
\end{equation}
Combining these two operations we get 
\begin{equation} \nonumber
    \st{4}=U_{\sf det}U_{e}\st{3}=-i \, \frac{e^{i(\chi+\gamma+\pih)}\su+\sd}{\sqrt 2}\ot\ket{\I \, E_1} .
\end{equation}
A DC spin flipper represented by the operator
\begin{eqnarray*} 
    U_{\sf DC}=\ket{\down}\bra{\up}+\ket{\up}\bra{\down},
\end{eqnarray*}
flips the spin
\begin{eqnarray*} 
    \st{5}=U_{\sf DC}\ket{\psi_4}=-i\, \frac{\ket{\up}+e^{i(\chi+\gamma+\frac{\pi}{2})}\ket{\down}}{\sqrt 2}\ot\ket{\I \, E_1} .
\end{eqnarray*}
The spin phase shift $U_{s}(\alpha)$  operator
\begin{equation}
    U_{s}=\ket{\up}\bra{\up}+e^{i \alpha} \ket{\down}\bra{\down} , \nonumber
\end{equation}
generates the final state
\begin{equation}
    \ket{\psi_f}=-i \frac{\ket{\up}+e^{i(\alpha+\chi+\gamma+\frac{\pi}{2})}\ket{\down}}{\sqrt 2}\otimes\ket{\I \, E_1}.  \nonumber
\end{equation}
This $\ket{\psi_f}$ passes through a $\frac{\pi}{2}$ spin turner and then enters the polarization analyzer and the detector. This combination realizes a projective measurement $P^s(0)$ on the spin state which counts neutrons in the state $\frac{\ket{\up}+\ket{\down}}{\sqrt 2}$. Let $N(\alpha, \chi, \gamma)$ be the neutron count rate with spin, path and energy phase shifts set to $\alpha$, $\chi$ and $\gamma$. One can show that
\begin{equation} \nonumber
    N(\alpha, \chi, \gamma) \propto 1+\cos(\alpha+\chi+\gamma+\frac{\pi}{2}).
\end{equation} 
Note that $N(\alpha, \chi,\gamma)$ is proportional to $\|P^s(\alpha) P^p(\chi)P^e(\gamma)|\psi_{\sf GHZ} \rangle\|^2$. 
The reason is that the path phase shifter $U_p(\chi)$ together with the projection operation $P^p_{\sf BR}$ act like a projection operation $P^p(\chi)$ on the path subsystem, $U_{\sf det}$ together with the energy phase shift $U_e(\gamma)$ and DC spin flipper act like a projection operation on the energy subsystem $P^e(\gamma)$, and the spin phase shift $U_s(\alpha)$ together with $P^s(0)$ act like a projective measurement $P^s(\alpha)$ on the spin. Therefore, 
\begin{equation} \label{Eq: counting statistics_Hasegawa}
    N(\alpha, \chi, \gamma) \propto E[P^s(\alpha)P^p(\chi)P^e(\gamma)].
\end{equation}

To measure the Mermin witness $M$, using this perfect crystal interferometer, one proceeds as in Sec. \ref{Sec: Statistical Analysis}.

\ignore{
One can measure a particular expectation value in Mermin witness $M$ of  Eq.~\eqref{Eq: Mermin-Witness} such as $\{\sigma^s_{u(\alpha)}, \sigma^p_{v(\chi)}, \sigma^e_{w(\gamma)}\}$ where $\alpha, \chi, \gamma=0,\frac{\pi}{2}$.  By decomposing the Pauli matrices into projectors $\s^l_{w(\phi)}=P^l(\phi)-P^l(\phi+\pi)$ and using Eq. \eqref{Eq: counting statistics_Hasegawa} one gets 
\begin{eqnarray}\nonumber
    E(\alpha, \chi, \gamma)&=&\frac{\sum_{\mu_s, \mu_p, \mu_e } (-1)^{\mu_s+\mu_p+\mu_e} E_{\mu_s\mu_p \mu_e}}{\sum_{\mu_s, \mu_p, \mu_e } E_{\mu_s\mu_p \mu_e}}\\
    &=&\frac{\sum_{\mu_s, \mu_p, \mu_e } (-1)^{\mu_s+\mu_p+\mu_e} N_{\mu_s\mu_p \mu_e}}{\sum_{\mu_s, \mu_p, \mu_e } N_{\mu_s\mu_p \mu_e}}, \nonumber
\end{eqnarray}
where $E_{\mu_s \mu_p \mu_e}=E[P(\alpha+\mu_s \pi) P( \chi+\mu_p \pi) P( \gamma+\mu_e \pi)]$ and $N_{\mu_s \mu_p \mu_e}=N(\alpha+\mu_s \pi, \chi+\mu_p \pi, \gamma+\mu_e \pi)$ using the count rates proportional to expectation value of the product of projectors. To determine the expectation value of that context one needs eight measurements with different phase shift settings $\{N(\alpha+\mu_s \pi, \chi+\mu_p \pi, \gamma+\mu_e \pi)\}$ with $\mu_s, \mu_p, \mu_e =0,1$. 
}

\end{document}